\documentclass[]{aastex}
\usepackage{emulateapj5}
\usepackage{onecolfloat}
\usepackage{graphicx} 
\usepackage{fancyheadings} 
\usepackage{ulem}
\usepackage{rotating}
\usepackage{lscape}

\newcommand{\delv}{\Delta v}
\newcommand{\xfeii}{[X$_i$/Fe$^+$]}
\newcommand{\xhi}{[X$_i$/H$^0$]}
\newcommand{\fN}{$f_N$(\feii)}
\newcommand{\suii}{S$^+$} 
\newcommand{\hi}{H$^0$} 
\newcommand{\hii}{H$^+$} 
\newcommand{\nti}{N$^0$} 
\newcommand{\ntii}{N$^+$} 
\newcommand{\ari}{Ar$^0$} 
\newcommand{\cii}{C$^+$} 
\newcommand{\ciis}{C\,II$^*$}
\newcommand{\civ}{C$^{3+}$} 
\newcommand{\alii}{Al$^+$} 
\newcommand{\aliii}{Al$^{++}$} 
\newcommand{\siii}{Si$^+$} 
\newcommand{\siiii}{Si$^{++}$} 
\newcommand{\siiv}{Si$^{3+}$} 
\newcommand{\crii}{Cr$^+$} 
\newcommand{\feii}{Fe$^+$} 
\newcommand{\feiii}{Fe$^{++}$} 
\newcommand{\nkii}{Ni$^+$} 

\newcommand{\lya}{Ly$\alpha$ }

\newcommand{\cm}[1]{\, {\rm cm^{#1}}}
\newcommand{\N}[1]{{N({\rm #1})}}

\newcommand{\sci}[1]{{\rm \; \times \; 10^{#1}}}

\newcommand{\perd}{\;\;\; .}
\newcommand{\cmma}{\;\;\; ,}

\newcommand{\mkms}{{\rm \; km/s}}

\begin{document}

\twocolumn[%
\submitted{Accepted to the Astrophysical Journal: Feb 5, 2002}

\title{THE UCSD HIRES/KECK\,I DAMPED \lya ABUNDANCE 
DATABASE\altaffilmark{1}
III. An Empirical Study of Photoionization in the Damped \lya 
System Toward GB1759+7539}

\author{ JASON X. PROCHASKA\altaffilmark{2,3}}
\affil{The Observatories of the Carnegie Institute of Washington}
\affil{813 Santa Barbara St. Pasadena, CA 91101}
\email{xavier@ociw.edu}
\and
\author{ J. CHRISTOPHER HOWK}
\affil{Department of Physics and Astronomy;
Johns Hopkins University; \\
3400 North Charles St.; Baltimore, MD 21218}
\email{howk@pha.jhu.edu}
\and
\author{JOHN M. O'MEARA\altaffilmark{2}, DAVID TYTLER\altaffilmark{2},
 ARTHUR M. WOLFE\altaffilmark{2}, DAVID KIRKMAN\altaffilmark{2},
DAN LUBIN, \& NAO SUZUKI}
\affil{Department of Physics, and Center for Astrophysics and Space Sciences}
\affil{University of California, San Diego; 
C--0424; La Jolla, CA 92093}
\email{jomeara@ucsd.edu, dtytler@ucsd.edu, awolfe@ucsd.edu, dkirkman@ucsd.edu
dlubin@ucsd.edu, n1suzuki@ucsd.edu}

\begin{abstract} 

We investigate the ionization state of the damped \lya system at
$z=2.62$ toward GB1759+7539 through an analysis of ionic ratios
sensitive to photoionization: 
\ari/\suii, \feiii/\feii, \ntii/\nti, \aliii/\alii. 
Approximately half of the metals arise in a mostly neutral velocity
component with H\,I/H~$> 0.9$, based on
\feiii/\feii $\,< 0.013$.
In contrast, the remaining half exhibits \feiii/\feii~$\approx 0.3$
indicative of a partially ionized medium with H\,I/H~$\approx 0.5$.
These conclusions are supported by the observed
\ntii/\nti, \aliii/\alii, and \ari/\siii\ ratios.

We assess ionization corrections for the observed column densities
through photoionization models derived from the CLOUDY software package.
In the neutral gas, the ionization corrections are negligible
except for \ari.  However for the partially ionized gas,
element abundance ratios differ from the ionic ratios by
0.1 -- 0.3~dex for (\siii, \suii, \nkii, \alii)/\feii\ ratios and more
for (\nti, \ari)/\feii.  
Independent of the shape of the photoionizing spectrum and assumptions
on the number of ionization phases,
these ionization corrections have minimal impact $(\lesssim 0.1$~dex)
on the total metallicity inferred for this damped \lya system.  
Measurements on the relative elemental
abundances of the partially ionized gas, however, have a greater than 
$\approx$0.15~dex uncertainty which hides the effects of 
nucleosynthesis and dust depletion.

We caution the reader that this damped system is unusual for a number
of reasons (e.g.\ a very low \ari/\suii\ ratio) and 
we believe its ionization properties are special but not unique.
Nevertheless, it clearly shows the value of examining photoionization
diagnostics like \feiii/\feii\ in a larger sample of systems. 

\end{abstract}

]

\pagestyle{fancyplain}
\lhead[\fancyplain{}{\thepage}]{\fancyplain{}{PROCHASKA ET AL.}}
\rhead[\fancyplain{}{THE UCSD HIRES/KECK\,I DAMPED \lya ABUNDANCE 
DATABASE III.}]{\fancyplain{}{\thepage}}
\setlength{\headrulewidth=0pt}
\cfoot{}

\altaffiltext{1}{http://kingpin.ucsd.edu/$\sim$hiresdla} 
\altaffiltext{2}{Visiting Astronomer, W.M. Keck Telescope.
The Keck Observatory is a joint facility of the University
of California and the California Institute of Technology.}
\altaffiltext{3}{Hubble fellow}

\section{INTRODUCTION}

Through echelle optical spectroscopy
of the damped \lya systems -- quasar absorption line systems with
$\N{HI} \geq 2 \sci{20} \cm{-2}$ -- one studies
the chemical abundances of high redshift galaxies in a manner
analogous with UV spectroscopy of the Galactic ISM. 
These observations yield ionic column density measurements of a series
low-ions\footnote{The term low-ion refers to the dominant ionic state
of a given element in an HI region.} including Si$^+$, Fe$^+$, Ni$^+$, 
Zn$^+$, S$^+$, and N$^0$ \citep[e.g.][]{lu96,pro99,molaro00,pro01}, often
at a greater accuracy than measurements of the Galactic ISM.
By studying these chemical abundances, one gains insight into
the nucleosynthetic enrichment,
dust properties, and metallicity of the damped \lya systems and
thereby the high redshift protogalactic population \citep[e.g.][]{pw02}.

To date, most damped \lya abundance studies have proceeded under the 
assumption that the gas has a very low ionization fraction, 
\begin{equation}
x \equiv {\rm \frac{H^+}{H^0 + H^+}} \ll 1 \perd
\end{equation}
In this case, the low-ion species
represent the dominant ionization states and ionization corrections for
elemental abundances are small, e.g.\ $\N{Si} \simeq \N{Si^+}$.  
There is a good theoretical basis for adopting this assumption.
\cite{pro96} performed a series of radiative transfer calculations and
predicted $x < 10\%$ for systems with $\N{HI} > 10^{20} \cm{-2}$.
Furthermore, the authors compared observed Al$^{++}$/Al$^+$
and Si$^{3+}$/Si$^+$ ratios for the damped \lya systems
against a series of photoionization models
derived from the CLOUDY software package 
\citep[v. 95;][]{ferland01} and concluded it has an ionization
fraction $x < 50 \%$.  \cite{viegas95} reached similar conclusions for
systems with $\N{HI} > 10^{21} \cm{-2}$ but warned that ionization corrections
could be important in systems with $\N{HI} \approx 10^{20} \cm{-2}$.
More recently, \cite{howk99} argued that photoionization
could imply significant corrections for some ionic species in the damped
\lya systems, particularly if there were a significant local source of
ionization (e.g. recent star formation).  
\cite{vladilo01} have extended this analysis
further and concluded that ionization 
corrections are a concern for only a few of the elements (e.g.\ Al) observed
in the damped systems. 
Finally, \cite{izotov01} have presented a specific model for the damped \lya
systems which suggests large ionization corrections for a number of
elements.  This model predicts O$^0$/\siii\ ratios, however, which are much
lower than those typically observed in the damped \lya systems. 

All of these studies have presented theoretical arguments which require
idealized physical conditions (e.g. constant volume density, plane-parallel
geometry), unknown quantities (e.g. shape, flux of the ionizing radiation),
and uncertain physics (e.g. dielectric recombination rates).
To overcome many of these uncertainties, one can empirically determine the
ionization state of the gas by measuring adjacent
ions of several elements (e.g.\ \feiii/\feii).
Observationally, this requires an analysis of transitions like
Si~III $\lambda 1206$, Fe~III $\lambda 1122$, and N~II $\lambda 1083$
all of which arise in the \lya forest and typically at observed wavelength
$\lambda < 4000$\AA.  Unfortunately, we have not tended 
to acquire these observations because of the limited wavelength coverage of the 
HIRES spectrograph in one setting and its 
poor sensitivity below 4000~\AA.  Therefore, the large damped \lya databases 
provided by the Keck community \citep{lu96,pro99,ptt99,pro01} 
have not directly addressed ionization corrections for
the damped \lya abundances.
Similarly, other studies in the literature \citep[e.g.][]{fan94}
have only focused on ratios like \civ/\cii\ which provide a less precise evaluation
of photoionization.

In this paper, we examine the photoionization of a single damped \lya system
through observations of a series of ions sensitive to the ionization
state of the gas.
We present a case study of the $z_{abs} = 2.62$ damped
\lya system toward GB1759+7539 ($z_{em} = 3.05$, V$\approx$17).  
This sightline has been
previously analysed by \cite{pro99}, \cite{outram99}, and \cite{pro01},
but none of these studies focused on the ionization state of the damped 
\lya system. 
Here, we present observations of a series of metal-line
transitions which indicate this damped \lya system is partially ionized.
Before proceeding, we emphasize that this damped \lya system was selected
on the basis that it exhibits several transitions suggestive
of photoionization (e.g. strong Si~III 1206 absorption).  
Therefore, {\it the system may not
be representative of the damped \lya population}.
In fact, there are at least four peculiar characteristics of this damped system:

\begin{enumerate}

\item It is the only known damped system with a significantly sub-solar 
Ar$^0$/Si$^+$ ratio \citep{pw02}.

\item It is one of the few systems where the low-ions
do not uniformly track one another in velocity space. 

\item It is one of the few systems where the Al~III profile shows
significant departures from the low-ion profiles \citep{wp00a}.

\item It is one of the few systems where the C~II$^*$ profile
does not closely trace the low-ion profiles \citep{wol01}.

\end{enumerate}
Interestingly, these characteristics are in much better general
agreement with gas probed in the Galactic ISM than with other damped systems
\citep[e.g.][]{sav96}.
Irrespective of the peculiar nature of the damped system toward
GB1759+7539, our analysis emphasizes the importance of 
observations of Si~III, Fe~III, N~II, Ar~I and other transitions for 
assessing the impact of photoionization on the chemical abundances one derives
from low-ion transitions.

\begin{table}[ht]\footnotesize
\begin{center}
\caption{{\sc ATOMIC DATA \label{tab:fosc}}}
\begin{tabular}{lccc}
\tableline
\tableline
Transition &$\lambda$ &$f$ & Ref\\
\tableline
 SIII 1012 & 1012.5020 & 0.03550000 &  1  \\
  ArI 1048 & 1048.2199 & 0.26280000 &  3  \\
  ArI 1066 & 1066.6600 & 0.06747000 &  3  \\
  NII 1083 & 1083.9900 & 0.10310000 &  1  \\
FeIII 1122 & 1122.5260 & 0.16200000 &  3  \\
   NI 1134a& 1134.1653 & 0.01342000 &  1  \\
   NI 1134b& 1134.4149 & 0.02683000 &  1  \\
   NI 1134c& 1134.9803 & 0.04023000 &  1  \\
 FeII 1143 & 1143.2260 & 0.01770000 &  2  \\
 SiII 1193 & 1193.2897 & 0.49910000 &  1  \\
   NI 1199 & 1199.5496 & 0.13280000 &  1  \\
   NI 1200a& 1200.2233 & 0.08849000 &  1  \\
   NI 1200b& 1200.7098 & 0.04423000 &  1  \\
SiIII 1206 & 1206.5000 & 1.66000000 &  1  \\
  SII 1250 & 1250.5840 & 0.00545300 &  1  \\
  CII 1334 & 1334.5323 & 0.12780000 &  1  \\
 CII* 1335 & 1335.7077 & 0.11490000 &  1  \\
 NiII 1370 & 1370.1310 & 0.07690000 &  5  \\
 SiIV 1393 & 1393.7550 & 0.52800000 &  1  \\
 SiIV 1402 & 1402.7700 & 0.26200000 &  1  \\
 SiII 1526 & 1526.7066 & 0.12700000 &  9  \\
  CIV 1548 & 1548.1950 & 0.19080000 &  1  \\
  CIV 1550 & 1550.7700 & 0.09522000 &  1  \\
 FeII 1608 & 1608.4511 & 0.05800000 & 17  \\
 FeII 1611 & 1611.2005 & 0.00136000 & 16  \\
 AlII 1670 & 1670.7874 & 1.88000000 &  1  \\
 NiII 1709 & 1709.6042 & 0.03240000 &  7  \\
 NiII 1741 & 1741.5531 & 0.04270000 &  7  \\
 SiII 1808 & 1808.0130 & 0.00218000 & 11  \\
AlIII 1854 & 1854.7164 & 0.53900000 &  1  \\
 ZnII 2026 & 2026.1360 & 0.48900000 & 13  \\
 CrII 2066 & 2066.1610 & 0.05150000 & 13  \\
\tableline
\end{tabular}
\end{center}
\tablerefs{Key to References -- 1:
\cite{morton91}; \\
2: \cite{howk00}; 3: \cite{morton02}; \\
5: \cite{fedchak99}; 7: \cite{fedchak00}; \\
9: \cite{schect98}; 10: \cite{bergs96}; \\
11: \cite{bergs93}; 13: \cite{bergs93b}; \\
17: \cite{bergs96b}}
\tablecomments{See http://kingpin.ucsd.edu/$\sim$hiresdla for the
most \\ current and complete version of this table.}
\end{table}
 
\begin{figure*}[ht]
\begin{center}
\includegraphics[height=6.0in, width=5.0in]{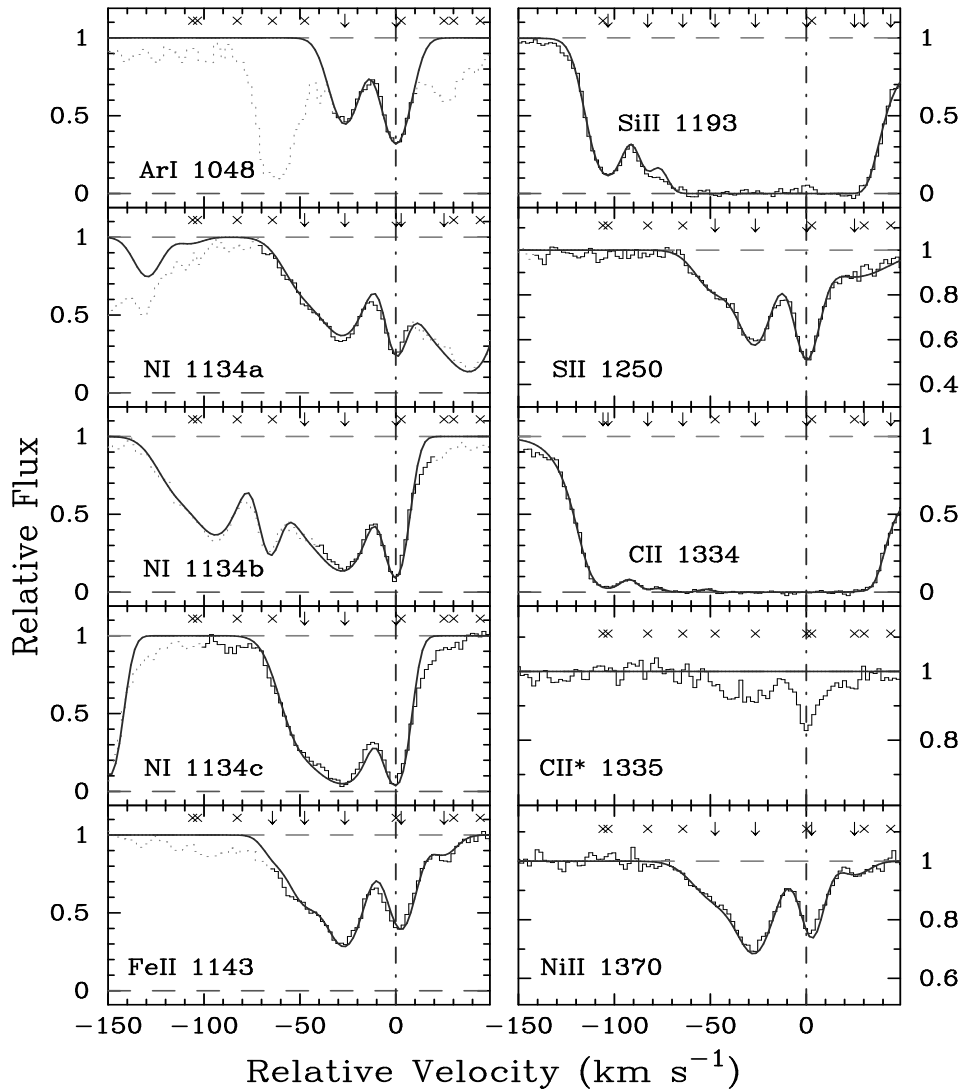}
\caption{Velocity profiles of low-ion transitions from the 
$z_{abs}=2.62$ damped \lya system toward GB1759+7539.  In the figure,
$v=0 \mkms$ corresponds to $z=2.62561$.
Most of the transitions are overplotted with a profile fit derived with the
VPFIT software package.  
The marks (arrows and x's) at the top of each panel indicate
all of the velocity components included in the fits. 
The arrows designate the components included in the analysis of a specific ion
whereas the x's have been excluded.
Known blends have been dotted out in the figure.
}
\label{fig:low}
\end{center}
\end{figure*}

\break

\section{IONIC COLUMN DENSITIES}
\label{sec-colm}

The quasar GB1759+7539 has been observed by at least three groups 
\citep{pro99,outram99,pro01} with the
HIRES spectrograph \citep{vogt94} on the Keck~I telescope.  
We have combined these data to produce a spectrum with 
nearly continuous wavelength coverage from 
$\lambda = 3400 - 7590$\AA.  
The \cite{pro99} and \cite{pro01} data sets were combined without
coadding because the spectral coverage did not overlap.  Therefore,
the resolution of the original observations was preserved at 
6~km/s and 8~km/s respectively.  Finally, we included the data 
\cite{outram99} to fill in the inter-order
gaps longward of $\approx 5000$~\AA.
In terms of the damped \lya system at
$z=2.625$ ($\log \N{HI} = 20.761 \pm 0.007$; Outram et al.\ 1999), 
the observations present measurements of transitions ranging
from S~III 1012 to Cr~II 2066.  

\begin{figure*}[ht]
\begin{center}
\includegraphics[height=6.0in, width=5.0in]{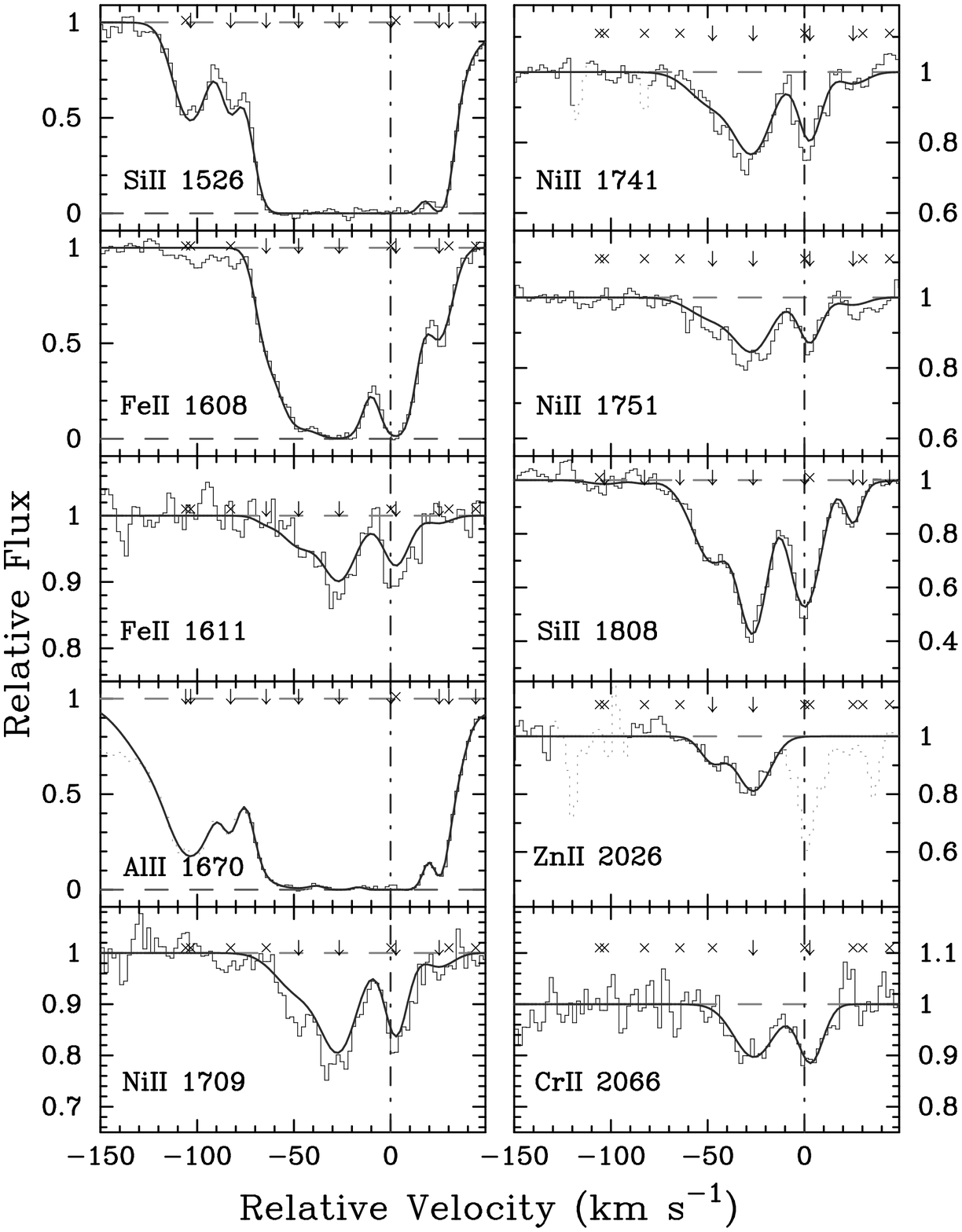}
\end{center}
\begin{center}
Fig1 -- cont
\end{center}
\end{figure*}

In the following sub-sections we derive
ionic column densities for the majority of transitions identified for
this damped system.  We have performed a line-profile fitting analysis
using the VPFIT software 
package\footnote{http://www.ast.cam.ac.uk/$\sim$rfc/vpfit.html} 
kindly provided by R. Carswell and J. Webb.
This least-squares routine minimizes a $\chi^2$ matrix
of multiple component Voigt profile fits to observed line profiles.  Each
velocity component is described by three parameters -- column density, 
redshift, and Doppler parameter ($b$-value) -- which can selectively be
'tied' together from ion to ion.  
It should be noted that line blending can lead to significant uncertainties
in the Doppler parameter and sometimes redshift of various components.
On the other hand, our experience is that the column densities remain
reasonably robust for unsaturated profiles.  
It is important to note that we include solutions for a number of heavily
saturated profiles.  These components are easily identified by their very
large reported uncertainties and they have not been included in the
analysis.  Furthermore, some velocity components have large uncertainties
because of a degeneracy in the solution (e.g.\ two velocity components which
nearly overlap).  We have taken care to insure that these line blending
effects have not influenced our photoionization analysis. 
Finally, Table~\ref{tab:fosc} presents
the atomic data used for the metal-line transitions analysed in this
paper.

\begin{table*}
\begin{center}
\caption{ {\sc
IONIC COLUMN DENSITIES: Low-Ions \label{tab:low}}}
\begin{tabular}{rcccrrrcr}
\tableline
\tableline
ID & $z_{abs}$ & $\sigma(z)$ & $v_{rel}$\tablenotemark{a}
& Ion & $\log N$ & $\sigma(N)$ & $b$& $\sigma(b)$ \\ 
 & & ($10^{-5}$) & (km/s) & & ($\cm{-2}$)  & & (km/s) \\
\tableline
 1&2.624329& 4.1&$-105$&C$^+$&13.747& 0.514&20.41&12.68\\  
&&&&Al$^+$&12.677& 0.093&29.23& 5.28\\  
 2&2.624360& 0.3&$-103$&Si$^+$&13.452& 0.010&10.67& 0.32\\  
&&&&C$^+$&14.160& 0.237&11.24& 2.78\\  
&&&&Al$^+$&12.428& 0.164&11.48& 1.65\\  
 3&2.624610& 0.4&$ -82$&Si$^+$&13.164& 0.042& 5.84& 0.65\\  
&&&&C$^+$&14.042& 0.082& 6.37& 1.68\\  
&&&&Al$^+$&12.008& 0.093& 4.94& 0.77\\  
 4&2.624832& 1.3&$ -64$&Si$^+$&13.505& 0.365& 8.76& 1.55\\  
&&&&Fe$^+$&13.453& 0.136& 6.32& 1.04\\  
&&&&C$^+$&14.794& 1.180& 7.21& 5.14\\  
&&&&Al$^+$&12.383& 0.409& 7.77& 1.82\\  
 5&2.625035& 0.8&$ -47$&Si$^+$&14.902& 0.050&12.40& 1.69\\  
&&&&Fe$^+$&14.331& 0.039&11.03& 1.12\\  
&&&&N$^0$&14.401& 0.038&13.72& 1.19\\  
&&&&Al$^+$&13.196& 0.119&14.56& 4.87\\  
&&&&Zn$^+$&11.732& 0.153& 7.69& 3.42\\  
&&&&Ni$^+$&13.149& 0.040&15.48& 1.47\\  
&&&&S$^+$&14.421& 0.065&12.72& 2.42\\  
 6&2.625289& 0.3&$ -26$&Si$^+$&15.129& 0.024& 8.75& 0.45\\  
&&&&Fe$^+$&14.700& 0.016&11.38& 0.38\\  
&&&&C$^+$&16.212& 8.658&11.10&23.77\\  
&&&&N$^0$&14.789& 0.017&14.08& 0.58\\  
&&&&Al$^+$&13.537& 1.752& 5.32& 4.91\\  
&&&&Zn$^+$&12.221& 0.060&11.06& 2.04\\  
&&&&Ni$^+$&13.463& 0.019&12.19& 0.59\\  
&&&&Cr$^+$&12.991& 0.094&13.78& 3.83\\  
&&&&S$^+$&14.786& 0.035&10.78& 0.79\\  
&&&&Ar$^0$&13.307& 0.028& 9.26& 1.04\\  
 7&2.625612& 0.1&+0&Si$^+$&15.057& 0.014& 9.77& 0.37\\  
&&&&C$^+$&17.831& 1.661& 8.07& 4.93\\  
&&&&N$^0$&14.627& 0.018& 5.93& 0.26\\  
&&&&Al$^+$&14.132& 0.622& 9.25& 2.24\\  
&&&&S$^+$&14.740& 0.126& 7.62& 1.34\\  
&&&&Ar$^0$&13.432& 0.021& 8.49& 0.67\\  
 8&2.625644& 0.2&+3&Fe$^+$&14.473& 0.010& 8.46& 0.21\\  
&&&&Ni$^+$&13.208& 0.019& 7.08& 0.51\\  
&&&&Cr$^+$&12.826& 0.110& 7.89& 2.73\\  
 9&2.625914& 0.3&+25&Si$^+$&14.216& 0.067& 4.29& 0.41\\  
&&&&Fe$^+$&13.602& 0.019& 7.80& 0.47\\  
&&&&Al$^+$&12.501& 0.157& 3.48& 1.15\\  
&&&&Ni$^+$&12.498& 0.097& 9.41& 3.10\\  
&&&&S$^+$&14.479& 0.540&23.68&32.07\\  
10&2.625975& 8.1&+30&Si$^+$&13.240& 0.568& 8.87& 4.92\\  
&&&&C$^+$&14.161& 0.542& 7.22& 4.70\\  
&&&&Al$^+$&12.087& 0.562& 7.65& 4.22\\  
11&2.626142&19.9&+44&Si$^+$&12.784& 0.533&17.23& 9.25\\  
&&&&C$^+$&13.499& 1.107&14.30&16.84\\  
&&&&Al$^+$&11.560& 0.479&18.75&12.47\\  
\tableline
\end{tabular}
\end{center}
\tablenotetext{a}{Velocity relative to z=2.62561}
\tablecomments{Portions of the C\,II 1334 and Al\,II 1670 transitions are heavily saturated in
some velocity regions and column densities measured for these transitions are highly uncertain.}
\end{table*}

\subsection{Low-Ion Profiles}

Figure~\ref{fig:low} presents velocity profiles for the most important 
low-ion transitions in our analysis.  
The velocity profiles are comprised of two main
features at $v \approx 0$ and $-25$ km/s where $v = 0 \mkms$ corresponds
to the redshift $z = 2.62561$.  Contrary to the majority
of damped \lya systems, 
it is obvious from the figure that the relative abundances
of these two components are not constant from ion to ion.  
The differences are less severe than the visual impression, but are greater 
than 0.1~dex (i.e.\ $> 3\sigma$)
in several cases.  This is a rare occurrence in the damped
\lya systems \citep[see][]{pro01} and it raises the likelihood that this system
is peculiar.

\begin{table*}[ht]\footnotesize
\begin{center}
\caption{ {\sc
IONIC COLUMN DENSITIES: Intermediate-Ions \label{tab:inter}}}
\begin{tabular}{rcccrrrcr}
\tableline
\tableline
ID & $z_{abs}$\tablenotemark{a} & $\sigma(z)$ & $v_{rel}$\tablenotemark{b}
& Ion & $\log N$ & $\sigma(N)$ & $b$\tablenotemark{c}& $\sigma(b)$ \\ 
 & & ($10^{-5}$) & (km/s) & & ($\cm{-2}$)  & & (km/s) \\
\tableline
 1&2.624360& 0.0&$-103$&Al$^{++}$&12.264& 0.049&11.64& 1.07\\  
 2&2.624832& 0.0&$ -64$&Al$^{++}$&12.128& 0.145&30.72&13.42\\  
&&&&N$^+$&13.909& 0.162&30.72& 0.00\\  
&&&&Fe$^{++}$&13.909& 0.137&30.72& 0.00\\  
 3&2.625035& 0.0&$ -47$&Al$^{++}$&12.922& 0.051&15.87& 1.32\\  
&&&&N$^+$&14.047& 0.098&15.87& 0.00\\  
&&&&Fe$^{++}$&13.919& 0.056&15.87& 0.00\\  
 4&2.625244& 0.7&$ -30$&Al$^{++}$&13.357& 0.022&16.50& 0.85\\  
&&&&N$^+$&14.764& 0.045&16.50& 0.00\\  
&&&&Fe$^{++}$&14.066& 0.066&16.50& 0.00\\  
 5&2.625612& 0.0&+  0&Al$^{++}$&12.629& 0.157& 5.27& 0.89\\  
&&&&N$^+$&13.838& 0.232& 5.27& 0.00\\  
 6&2.625697& 6.5&+7&Al$^{++}$&12.429& 0.285&10.79& 3.05\\  
&&&&N$^+$&13.931& 0.205& 6.81& 2.62\\  
 7&2.625914& 0.0&+ 25&Al$^{++}$&12.105& 0.058& 9.29& 0.70\\  
&&&&N$^+$&13.918& 0.034& 9.29& 0.00\\  
 8&2.626142& 0.0&+ 44&Al$^{++}$&11.710& 0.329&27.48&26.17\\  
&&&&N$^+$&13.473& 0.385&27.48& 0.00\\  
\tableline
\end{tabular}
\end{center}
\tablenotetext{a}{Components 1--3,5,8,9 were fixed  to have identical redshifts as components observed in  the low-ion profiles.}
\tablenotetext{b}{Velocity relative to z=2.62561}
\tablenotetext{c}{Components with identical $b$ values were tied together in the fit.  In this case, an error is listed for only one component.}
\end{table*}

Table~\ref{tab:low} presents the full VPFIT solution and errors for 
the low-ion profiles derived with the VPFIT software package.  
In Figure~\ref{fig:low}, 
most of the transitions are overplotted with the profile fit and
the marks (arrows and x's) at the top of each panel indicate
all of the velocity components included in the fits. 
The arrows designate the components included in the analysis of a specific ion
whereas the x's identify excluded components.
In general, all of the low-ions were fit with 3--7 components with tied
redshifts.  In contrast to our previous experience with
VPFIT solutions of damped \lya profiles \citep[e.g.][]{pro96}, 
we found a significant improvement
in $\chi^2$ if the Doppler parameters were left untied from ion to ion.
Even more unusual, however, is an observed
offset in the $v \approx 0 \mkms$ component between the 
Fe~II, Ni~II, and Cr~II transitions and the other low-ion transitions.  
We found
a significant improvement in $\chi^2$ if we fit the former transitions
with a velocity component offset by $+3 \mkms$. 
We have experimented with our VPFIT solution and have been able to shift
the centroids of the Ni~II and Fe~II components by $\approx 1 \mkms$
closer to the other ions
by introducing another velocity component at $v \approx 10 \mkms$.  
The resulting fit, however, is not a significant improvement in $\chi^2$
over the present solution.  We are concerned with this
offset, in particular its implications for analyses which are
extremely sensitive to such effects \citep[e.g.][]{webb01}.
It would have only a small impact on the current analysis,
however, and we proceed under the assumption that
all of the gas near $v = 0 \mkms$ has the same velocity.

\begin{figure}[hb]
\begin{center}
\includegraphics[height=4.6in, width=3.5in]{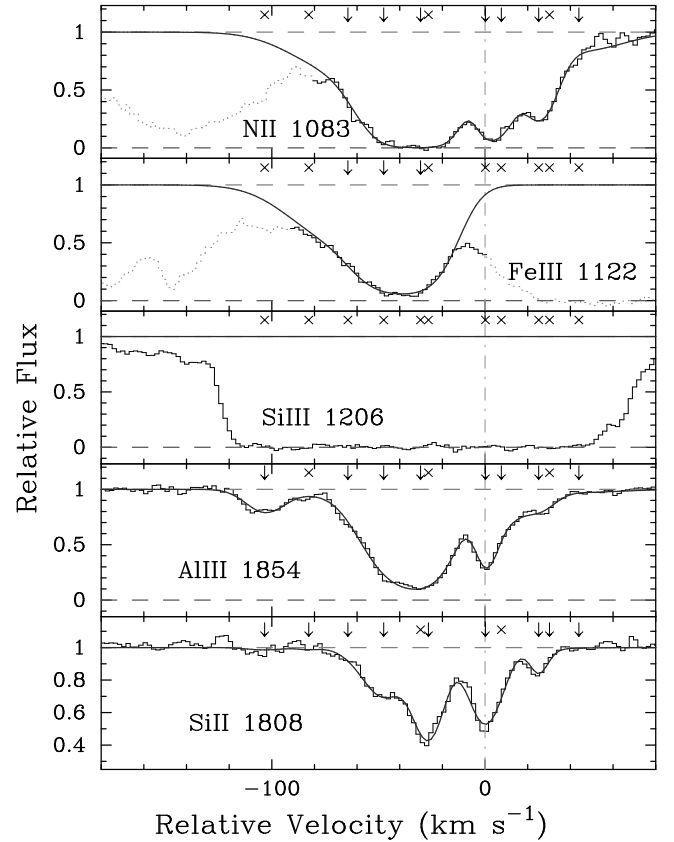}
\caption{ 
Same as Figure~\ref{fig:low} but for the intermediate-ion transitions.
For comparison, we also include the Si\,II 1808 low-ion profile.
}
\label{fig:inter}
\end{center}
\end{figure}

\subsection{Intermediate-Ion Profiles}

We present a series of intermediate-ion profiles in Figure~\ref{fig:inter}
which includes transitions of \feiii, \siiii, \aliii, and \ntii. 
To facilitate comparison with the low-ion transitions, we also plot the
Si~II 1808 transition.  
With the exception of 
Al~III 1854, the profiles are located within the \lya forest and may
be contaminated by coincident \lya clouds (we are confident that the dotted
regions in Figure~\ref{fig:inter} are blends).  For this reason, one may
consider these column density measurements as upper limits except
in the case where the profiles are heavily saturated and a precise
estimate is impossible.
We have fit these transitions with the VPFIT software package 
except Si~III 1206 which we found to be too saturated to include 
in the analysis.  If this ion has a similar component structure
to the other intermediate-ions, then
the majority of components have $\N{Si^{++}} > 10^{14} \cm{-2}$. 
We adopt this limit throughout. 

\begin{table*}\footnotesize
\begin{center}
\caption{ {\sc
IONIC COLUMN DENSITIES: High-Ions \label{tab:high}}}
\begin{tabular}{rcccrrrcr}
\tableline
\tableline
ID & $z_{abs}$ & $\sigma(z)$ & $v_{rel}$
& Ion & $\log N$ & $\sigma(N)$ & $b$& $\sigma(b)$ \\ 
 & & ($10^{-5}$) & (km/s) & & ($\cm{-2}$)  & & (km/s) \\
\tableline
 1&2.624064& 6.6&$-127$&C$^{3+}$&13.082& 0.097&27.54& 5.58\\  
 2&2.624293& 0.3&$-108$&C$^{3+}$&13.335& 0.036& 5.27& 0.50\\  
&&&&Si$^{3+}$&13.078& 0.043& 7.99& 0.32\\  
 3&2.624515& 0.4&$ -90$&C$^{3+}$&13.997& 0.020& 8.16& 0.48\\  
&&&&Si$^{3+}$&13.062& 0.474& 9.74& 2.39\\  
 4&2.624769& 4.4&$ -69$&C$^{3+}$&13.344& 0.287& 9.51& 5.39\\  
&&&&Si$^{3+}$&12.671& 2.049&19.34&28.56\\  
 5&2.624940& 4.1&$ -55$&C$^{3+}$&13.426& 1.201& 6.74& 4.60\\  
&&&&Si$^{3+}$&13.224& 0.302&13.23&10.95\\  
 6&2.625097& 2.7&$ -42$&C$^{3+}$&14.084& 0.462&11.57&11.85\\  
&&&&Si$^{3+}$&13.721& 0.156& 9.49& 1.45\\  
 7&2.625291& 2.4&$ -26$&C$^{3+}$&13.467& 1.444& 6.61& 8.54\\  
&&&&Si$^{3+}$&13.419& 0.297& 9.62& 4.71\\  
 8&2.625417&11.5&$ -15$&C$^{3+}$&13.374& 0.943& 7.92& 8.63\\  
&&&&Si$^{3+}$&13.020& 1.713&13.49&47.17\\  
 9&2.625601& 5.8&$   0$&C$^{3+}$&13.390& 0.296&10.82& 5.37\\  
&&&&Si$^{3+}$&12.830& 1.236& 9.37& 4.35\\  
10&2.625880& 2.9&+22&C$^{3+}$&13.199& 0.117& 9.86& 2.53\\  
&&&&Si$^{3+}$&12.847& 0.199&12.26& 5.16\\  
11&2.626038& 2.8&+35&C$^{3+}$&12.676& 0.341& 7.13& 3.59\\  
&&&&Si$^{3+}$&12.508& 0.336& 9.39& 4.50\\  
12&2.626206& 1.0&+49&C$^{3+}$&12.603& 0.106& 5.01& 1.38\\  
&&&&Si$^{3+}$&12.341& 0.137& 5.09& 1.02\\  
13&2.626430& 0.5&+68&C$^{3+}$&12.847& 0.047&16.30& 2.08\\  
&&&&Si$^{3+}$&12.236& 0.065&18.97& 3.30\\  
\tableline
\end{tabular}
\end{center}
\end{table*}

Similar to Figure~\ref{fig:low}, the marks at the top of each panel indicate
the velocity centroids of the components included in the analysis with the
arrows identifying the specific components included for a given ion.
As described below, it is an analysis of these
ions along with the low-ions which places the tightest constraints on the
ionization state of the damped \lya system.  
Therefore, we have attempted to fit the intermediate-ions with the
line-profile solution obtained from the low-ion transitions, i.e.,
the components are placed at the same velocities as the low-ion solution.
We achieved a satisfactory fit for the intermediate-ion profiles except
at $v \approx -26 \mkms$ and $v \approx +8 \mkms$.
At $v \approx -26 \mkms$, the $\chi^2$
of the fit was significantly improved if the velocity component was offset
by $-4 \mkms$ from the respective low-ion feature.  
We contend this offset is a physical effect and its implications on our
photoionization modeling are discussed below.
Meanwhile, we had to introduce a new component at $v \approx 8 \mkms$
to account for significant absorption at that velocity in the intermediate-ions.
Table~\ref{tab:inter}
presents the full solution for the intermediate-ion profiles.

\begin{figure}[hb]
\begin{center}
\includegraphics[height=4.0in, width=3.2in]{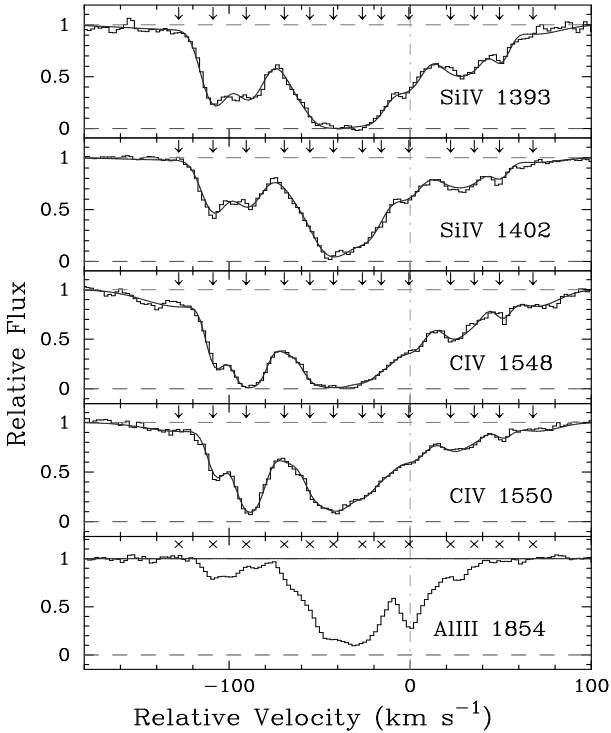}
\caption{
Same as Figure~\ref{fig:low} but for the high-ion transitions.
For comparison, we also include the Al\,III 1854 intermediate-ion profile.
}
\label{fig:high}
\end{center}
\end{figure}

\begin{table*}\footnotesize
\begin{center}
\caption{ {\sc
IONIC COLUMN DENSITIES: $v=-330$km/s Component\label{tab:sngl}}}
\begin{tabular}{rcccrrrcr}
\tableline
\tableline
Ion & $z_{abs}$ & $\sigma(z)$ & $v_{rel}$
& $\log N$ & $\sigma(N)$ & $b$& $\sigma(b)$ \\ 
 & & ($10^{-5}$) & (km/s) & ($\cm{-2}$)  & & (km/s) \\
\tableline
C$^+$&2.621580& 0.6&$-332$&12.775& 0.035& 4.36& 0.96\\  
C$^{3+}$&2.621484&12.1&$-340$&13.030& 0.876& 6.62& 5.17\\  
C$^{3+}$&2.621580& 3.0&$-332$&13.455& 0.337& 5.45& 1.33\\  
Si$^+$&2.621590& 0.5&$-332$&12.387& 0.031& 4.64& 0.85\\  
Si$^{++}$&2.621584& 0.3&$-332$&12.526& 0.028& 5.00& 0.44\\  
Si$^{3+}$&2.621570& 0.1&$-333$&12.956& 0.016& 3.54& 0.15\\  
Al$^{++}$&2.621571& 0.8&$-333$&11.537& 0.112& 5.39& 2.39\\  
Al$^+$ & ... & ... & $-333$ & $<10.8$ \\
\tableline
\end{tabular}
\end{center}
\end{table*}

\subsection{High-Ion Profiles}

In Figure~\ref{fig:high}, we present the profiles and VPFIT solution for the
high-ion Si~IV and C~IV transitions.
We have tied the redshifts of the velocity components for the two ions
but have allowed the $b$-values to vary.  We chose to fit the profiles
independently of the low and intermediate-ions -- even though several of the
velocity components have similar redshifts -- because we
believe the high-ion gas is spatially separated from the low-ion gas
\citep{wp00a}. 
For comparison, we plot an intermediate-ion Al\,III transition 
to emphasize its likeness to the Si~IV profiles.
The parameters of the solution are listed in Table~\ref{tab:high}.

\begin{figure}[ht]
\begin{center}
\includegraphics[height=4.5in, width=3.5in]{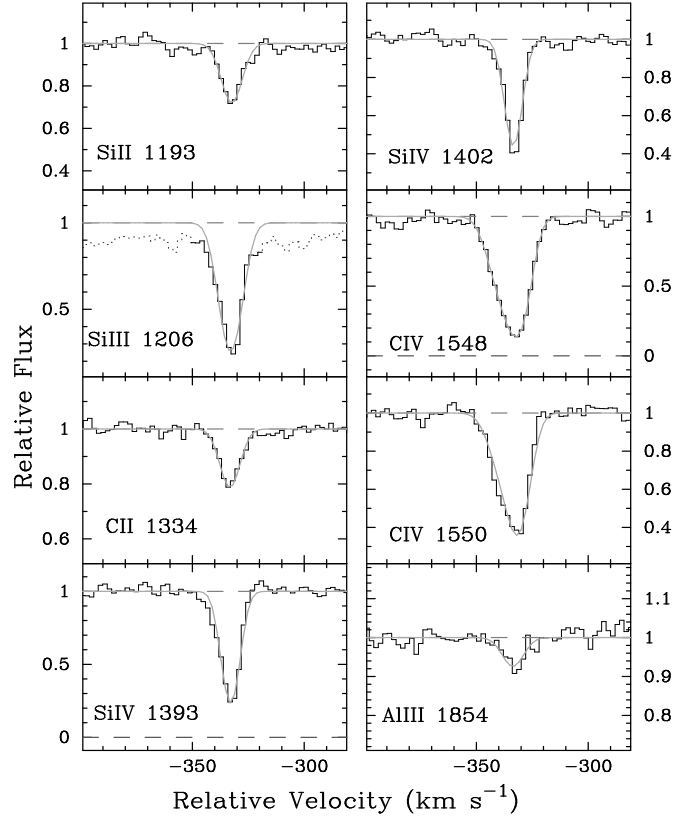}
\caption{
Velocity profiles of metal-line transitions from an absorption
system associated with the 
$z_{abs}=2.62$ damped \lya system toward GB1759+7539.  
This system has very simple kinematics, suggesting the multiple
ionic species arise in a single phase of gas.
Most of the transitions are overplotted with a profile fit derived with the
VPFIT software package (Table~\ref{tab:sngl}).  
Known blends have been dotted out in the figure.
}
\label{fig:sngl}
\end{center}
\end{figure}

\subsection{$v \approx -330 \mkms$ Feature}

Finally, we analysed a single absorption feature at $v=-330 \mkms$ 
which exhibits significant Si$^+$, \siiii, \siiv, C$^+$, C$^{3+}$, and Al$^{++}$
absorption (Figure~\ref{fig:sngl}).
We fit all of the profiles with a single component except 
\civ\ whose asymmetric shape indicates two components.
The centroids of all of the ions are at
nearly identical velocity including the stronger component of the C\,IV profiles.
The column densities for these ions are very accurately
measured (Table~\ref{tab:sngl}) and they provide a precise examination
of the ionization state of this gas.

\begin{table*}[ht]
\begin{center}
\caption{ {\sc SUMMARY TABLE \label{tab:summ}}}
\begin{tabular}{rrrrrrrrrr}
\tableline
\tableline
Ratio & $-333$& $-105$& $ -65$& $ -47$& $ -26$& $   0$& $  25$\\
\tableline
C$^{3+}$/C$^+$&$ 0.80$&$-0.97$&$<-1.67$&...&$<-2.95$&$<-4.45$&...\\  
Si$^{3+}$/Si$^+$&$ 0.56$&$-0.37$&$-0.83$&$-1.18$&$-1.71$&$-2.23$&$-1.37$\\  
Al$^{++}$/Al$^+$&$> 0.74$&$>-0.73$&$-0.26$&$<-0.49$&$<-0.29$&$-1.50$&$-0.40$\\  
N$^+$/N$^0$&...&...&$> 0.21$&$-0.35$&$-0.02$&$-0.79$&...\\  
Fe$^{++}$/Fe$^+$&...&...&$ 0.46$&$-0.41$&$-0.63$&$<-1.88$&...\\  
Si$^{++}$/Si$^+$&$ 0.12$&$> 0.15$&...&...&...&...&...\\  
Fe$^{++}$/Al$^{++}$&...&...&$ 1.78$&$ 1.00$&$ 0.71$&$<-0.04$&...\\  
Fe$^{++}$/N$^+$&...&...&$ 0.00$&$-0.13$&$-0.70$&$<-1.25$&...\\  
$[$Al$^{++}$/Si$^+$$]$ &$ 0.22$&$-0.12$&$-0.31$&$-0.91$&$-0.70$&$-1.36$&$-1.04$\\  
$[$Ar$^0$/S$^+$$]$ &...&...&...&...&$-0.80$&$-0.63$&...\\  
$[$Si$^+$/Fe$^+$$]$ &...&...&$-0.01$&$ 0.51$&$ 0.37$&$ 0.52$&$ 0.55$\\  
$[$Ni$^+$/Fe$^+$$]$ &...&...&...&$ 0.07$&$ 0.01$&$-0.02$&$ 0.15$\\  
$[$S$^+$/Fe$^+$$]$ &...&...&...&$ 0.39$&$ 0.39$&$ 0.57$&$ 1.18$\\  
$[$Cr$^+$/Fe$^+$$]$ &...&...&...&...&$ 0.12$&$ 0.18$&...\\  
$[$Zn$^+$/Fe$^+$$]$ &...&...&...&$ 0.23$&$ 0.35$&...&...\\  
$[$N$^0$/S$^+$$]$ &...&...&...&$-0.79$&$-0.77$&$-0.88$&...\\  
$[$N$^0$/Si$^+$$]$ &...&...&$<-0.21$&$-0.91$&$-0.75$&$-0.84$&...\\  
$[$Fe$^{++}$/Al$^{++}$$]$ &...&...&$ 0.77$&$-0.01$&$-0.30$&$<-1.05$&...\\  
$\%[N_{TOT}$(Si$^+$)] &$ 0.1 $&$ 0.8 $&$ 0.9 $&$22.5 $&$37.9 $&$32.2 $&$ 4.6 $\\  
$\%[N_{TOT}$(S$^+$)] &...&...&...&$15.3 $&$35.4 $&$31.9 $&$17.5 $\\  
$\%[N_{TOT}$(Fe$^+$)] &...&...&$ 2.6 $&$19.8 $&$46.4 $&$27.5 $&$ 3.7 $\\  
$\%[N_{TOT}$(Fe$^{++}$)] &...&...&$28.5 $&$29.2 $&$40.9 $&$< 1.4 $&...\\  
$\%[N_{TOT}$(Al$^{++}$)] &$ 0.8 $&$ 4.2 $&$ 3.1 $&$19.3 $&$52.5 $&$ 9.8 $&$ 2.9 $\\  
$\%[N_{TOT}$(N$^+$)] &...&...&$ 7.8 $&$10.7 $&$55.8 $&$ 6.6 $&$ 8.0 $\\  
$\%[N_{TOT}$(N$^0$)] &...&...&$< 3.7 $&$18.8 $&$45.9 $&$31.6 $&...\\  
\tableline
\end{tabular}
\end{center}
\end{table*}

\subsection{Summary}

Table~\ref{tab:summ} presents an observational summary of various
ionic ratios and column densities for the majority of velocity components
identified in the metal-line profiles.  The velocities listed at the top
of the Table designate 10~km/s bins where all of the gas within 5~km/s of
the centroid has been summed for the ionic column densities.  All
ratios are logarithmic values with the square bracket notation indicating
values relative to solar abundances, i.e.,
[X/Y]~$\equiv \log(\N{X}/\N{Y}) - \log({\rm X/Y})_\odot$.

\section{ANALYSIS OF THE GAS AT $v < -50 \mkms$}

In this section we consider various diagnostics on the ionization state
of the gas at $v < -50 \mkms$ along the sightline.  
It is important to keep in mind that the absorption at any velocity may
be the result of several components which simply overlap in velocity space.
These components might each have a unique set of physical characteristics
(e.g. metallicity, ionization state) and therefore comparisons of different
ions at a given velocity may be open to several interpretations.

\subsection{Gas at $v \approx -330 \mkms$}

As revealed by Figure~\ref{fig:sngl}, there is an absorption feature at
$v \approx -330 \mkms$ which exhibits significant absorption for many
ions: Si$^+$, Si$^{++}$, Si$^{3+}$, C$^{3+}$, C$^+$, and Al$^{++}$. 
The kinematic structure is very simple and we fit the profiles with a single
component (Table~\ref{tab:sngl}).  In fact,
the feature is so simple that we are confident the majority of
gas is spatially well-localized.
The \siiv, \civ, and
\aliii\ ions have significantly higher column densities than their low-ion 
counterparts indicating the gas is highly ionized.  
Significant \siiv\ and \civ\ column densities can arise from a number
of physical processes including collisional ionization and 
photoionization by high energy photons.
Because the observed 
Doppler parameters are small, we place a 2$\sigma$ 
upper limit on the temperature independently from the \cii\
and \siiv\ measurements: $T < 3 \sci{4}$~K.  
In turn,  this indicates the 
gas is not in collisional ionization equilibrium because
$T \gtrsim 4 \sci{4}$~K is required to produce significant amounts
of \civ\  \citep[c.f.][]{sutherland93}.
This naturally suggests the ionization may be dominated by photoionization.

Although this system is well separated by velocity from the 
damped \lya system, its characteristics are important for several reasons.
First, it is possible that this gas is representative of the gas
giving rise to the C~IV and Si~IV profiles presented in Figure~\ref{fig:high}.
Therefore, the gas responsible for the C~IV and Si~IV profiles makes
a negligible contribution to the low-ion column
densities -- and presumably $\N{HI}$ -- of the damped \lya system.
A similar conclusion has been inferred from the observed
differences in the low and high-ion profiles of many other damped \lya systems
\citep{lu96,wp00a}.  
We also note that the \siiii/\siiv\ and \aliii/\siiv\ ratios are sufficiently
small for the $v \approx -330 \mkms$ component that gas with its
ionization characteristics cannot reproduce both the high and 
intermediate-ion column densities at $v \approx 0 \mkms$.
If we attribute the C\,IV and Si\,IV profiles at $v>-50 \mkms$
in Figure~\ref{fig:high} to an 
ionized gas similar to the $v \approx -330 \mkms$ component, 
then we must introduce at least one additional phase of
gas for the intermediate-ions {\it independent of the low-ions.}
Therefore, if the intermediate and low-ions have distinct origins, this
indicates the presence of three unique ionization phases at a similar
velocity in this damped \lya system.

The absorption feature at $v \approx -330 \mkms$
is further important because it poses
a difficult challenge to ionization models.  
Because of its simple kinematics and low temperature, the gas 
is very likely in a single component with
a single set of physical conditions.  
Equilibrium collisional ionization models or more
exotic turbulent mixing layers \citep[e.g.][]{shull94} are inconsistent with
the temperature of the gas. 
If the gas is in equilibrium, therefore, then we
presume it has been photoionized.

With the range of ions and very accurate column densities observed, 
we expected to precisely constrain its ionization state. 
To this end, we performed a series of photoionization calculations using
the CLOUDY software package with both an 
extragalactic UV background spectrum \citep{haa96} and a series of
\cite{kurucz88} O-star spectra.  We varied
the HI column density of the plane-parallel 'cloud', 
assumed a range of metallicities,
and calculated the ionization state at a range of ionization 
parameters,

\begin{equation}
U \equiv \frac{\phi}{c n_H} \gtrsim 
(2 \sci{-5}) \, \frac{J_{912}/10^{-21.5}}{n_H/ \cm{-3}} \cmma
\label{eqn:U}
\end{equation}
where $n_H$ is the volume density of hydrogen, $\phi$ is the
surface flux of ionizing photons with $h \nu > 1$~Ryd, and $J_{912}$
is the intensity of the incident radiation at 1 Ryd.
The relation in Equation~\ref{eqn:U} approximately holds
for ionizing radiation fields which steeply decline at energies
greater than 1~Ryd.

To our surprise, we cannot identify any single-phase
(e.g.\ constant density) photoionization model which can
match the observed ratios of the Si ions 
(\siii\,:\,\siiii\,:\,\siiv = 1\,:\,1.3\,:\,4). 
The principal conflict is that models which predict 
$\N{Si^+} \approx \N{Si^{++}}$ always require $\N{Si^{++}} > \N{Si^{3+}}$
in stark contradiction with our observations.  This 
difference between the predicted
and observed ratios is generally greater than a factor of 10.
It holds independently of the shape or flux of the
input spectrum or any assumptions on the metallicity or H~I column density.
We also note that even if we allow the temperature to be a free parameter,
it is impossible to reproduce the Si ionic ratios through a
combination of collisional and photoionization. 
In conclusion, we contend {\it the gas cannot be 
described by a single ionization phase in equilibrium conditions.}  
While this could be a common occurrence in quasar absorption line
systems, it is very surprising for this feature because of its simple
kinematics.

Granted the failure of the single-phase equilibrium models,
we also considered the prospect that the gas
arises in two distinct phases with very different ionization states.  
If we combine a very neutral component with a highly ionized component,
we approach the observed ratios of Si and C but still fail at 
the $> 95\%$ c.l.  Furthermore, 
the best fitting model couples an extremely neutral gas with
an extremely photoionized gas in a mixture of $\approx 2:1$,
which would be unlikely given the
extremely simple kinematics of the observed metal-line profiles.  
It would take a contrived set of physical conditions to couple two-phases
of gas with such extreme ionization characteristics at identical velocities
and both with $T < 3 \sci{4}$~K. 
Finally, this two-phase model predicts an O\,I column density in 
conflict with our observed upper limit.
To this end, we have failed to appropriately model the ionization
state of this velocity component.  Although it has no
direct bearing on our damped \lya analysis because of its very
small metal content, we interpret this failure
as a stern warning for the challenges associated with making ionization
corrections for highly ionized gas. 

Our photoionization models do not accurately predict the ionization
structure of the $-330$ km/s complex; furthermore, collisional
ionization equilibrium models cannot provide the required ionization
since the upper limits on the kinetic temperature ($T < 3\times10^4$
K) are too small to allow collisional production of C\,IV and Si\,IV in
great quantities (Sutherland \& Dopita 1993).  It is possible that the
gas in this component is not in ionization equilibrium.
Non-equilibrium ionization can occur, for example, if the cooling time
of shock-heated gas is significantly longer than the recombination
time (Shapiro \& Moore 1976; Sutherland \& Dopita 1993).  Models of
time-dependent ionization of hot gas (e.g., gas hot enough to produce
significant amounts of the highly-ionized species observed in this
cloud) generically predict much larger ionization fractions of
highly-ionized species at the temperatures appropriate for this
component than do equilibrium models.  Thus another potentially valid
model for the $-330$ km/s gas is that this component was initially
shock-heated to high temperatures ($T > 10^5$ K) and has cooled
relatively quickly, with its ionization state lagging its temperature.
While the ionization fractions of \civ\ and \siiv\ can be large even to
temperatures as low as $10^4$ K in such scenarios (Sutherland \&
Dopita 1993), we note that models of non-equilibrium ionization
balance are dependent upon the initial conditions of the gas.  Thus it
is difficult to determine if the ionization structure of the $-330$
km/s absorption is consistent with this non-equilibrium scenario
without unfounded guesses at its initial temperature, density, and
abundance.

\subsection{Gas at $v \approx -105$ and $- 65 \mkms$}
\label{sub:intmgas}

The low column density 
gas with $v \approx -105$ and $-65 \mkms$ exhibits significantly 
different ionization
characteristics than the $v \approx -330 \mkms$ component. 
Although the ionic ratios (e.g.\ \siiii/\siii, \feiii/\feii, and \ntii/\nti)
all indicate the gas is predominantly ionized $(x \approx 1$), the
\civ\ and \siiv\ ions comprise a much smaller fraction of the elemental
column densities than the $v \approx -330 \mkms$ component.  

The properties of the gas at $v \approx -65, -105 \mkms$ may have important
implications for the intermediate-ion gas observed at $v > -50 \mkms$.
Consider the possibility that the low and intermediate-ion profiles
at $v > -50 \mkms$ arise from two ionization phases: (i) a neutral gas
phase and (ii) a highly ionized gas with properties similar to the
gas at $v \approx -65, -105 \mkms$.  In this case, we 
can estimate the contribution of this ionized phase to the total \feii\ column
density at $v > -50 \mkms$ from the following treatment.  
At $v \approx -65 \mkms$, we measure
a \feiii/\feii\ ratio of $+0.46$~dex.  If we assume the same
ratio applies at $v \approx -26 \mkms$, then the $\N{Fe^{++}}$ value at
$v \approx -26 \mkms$ implies $\N{Fe^+} = 10^{14} \cm{-2}$ which is
$\approx 20\%$ of the total $\N{Fe^+}$ value at that velocity.  
Therefore, an ionized gas component
could comprise roughly $20\%$ of the low-ion absorption observed
at $v \approx -26 \mkms$.  We develop this treatment to much greater
depth in the following section.

\section{ANALYSIS OF THE LOW-ION REGIONS AT $v > -50 \mkms$}

The three velocity components at 
$v \approx -47, -26$, and 0 km/s comprise nearly all of the
low-ion gas in this damped \lya system.  
In terms of the system's chemical abundances, therefore,
ionization corrections on the column densities derived 
from the low-ion profiles at these velocities are the most important.

\begin{figure*}[ht]
\begin{center}
\includegraphics[height=6.0in, width=4.5in,angle=-90]{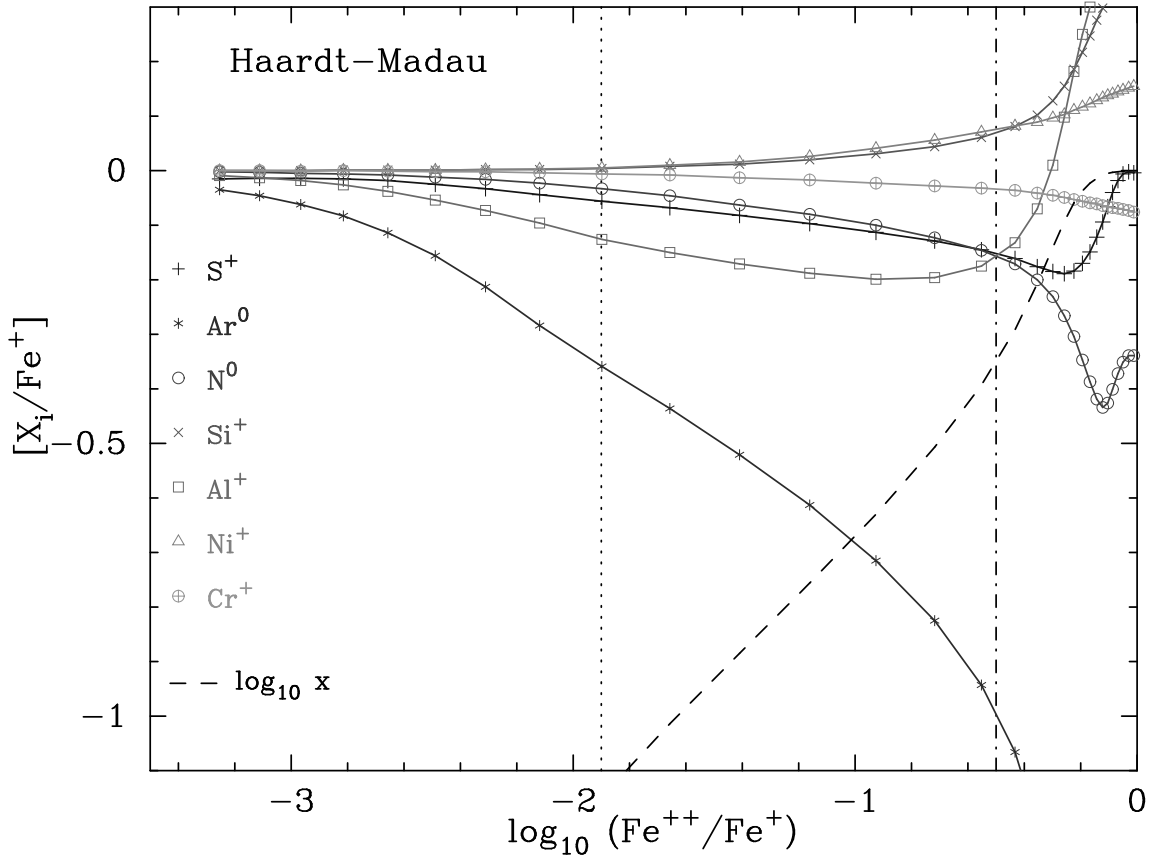}
\caption{
Predicted logarithmic abundance of low-ion X$_i$ 
relative to \feii\ relative to the intrinsic abundance of these two elements
(Equation~\ref{eqn:xfe}) against the \feiii/\feii\ ratio. 
Departures of \xfeii\ from zero 
indicate that photoionization
corrections are required to calculate accurate elemental abundance ratios
from the low-ion column densities independent of dust depletion.
Positive (negative) \xfeii\ values would imply overestimates (underestimates)
of [X/Fe] from the low-ion ratios.
The black dashed line plots the ionization fraction, $\log x$, on
the same logarithmic scale as \xfeii.
For this analysis, we have assumed the Haardt-Madau spectrum, 
$\N{HI} = 10^{20} \cm{-2}$, and [Fe/H] = --1.
All of these quantities were derived with the CLOUDY software package.
}
\label{fig:FeFe}
\end{center}
\end{figure*}

\subsection{$v \approx 0 \mkms$}

It is most instructive to begin with the $v \approx 0 \mkms$ component
where, initially,  
there are conflicting photoionization diagnostics.  
On the one hand, the \ari/\suii\ ratio is significantly sub-solar
which indicates the majority of Ar is ionized \citep[e.g.][]{sofia98}:
[\ari/\suii]~$\approx -0.7$~dex. 
Therefore, one might expect photoionization is
significantly affecting the chemical abundances 
inferred from all of the low-ion transitions.
On the other hand, we have placed a strict upper limit on the
\feiii\ column density which limits the level of photoionization of this
gas.  The adopted limit of $\log \N{Fe^{++}} < 12.6$~dex results from the 
low optical depth observed at $v \approx 0 \mkms$ in the Fe\,III 1122 profile
(Figure~\ref{fig:inter}).  We stress that the absorption present at 
$v \approx 0 \mkms$ is probably due to an intervening \lya cloud 
and therefore the $\N{Fe^{++}}$ value
is possibly lower than $10^{12} \cm{-2}$.  Even adopting our more conservative
upper limit, \feiii/\feii\ $< -1.9$~dex 
(Tables~\ref{tab:low},\ref{tab:inter}),
this indicates the gas has a small ionization fraction.  
Finally, the moderate strength charge-exchange reactions between
N and H which roughly couple the column densities of \nti, \ntii, \hi, and \hii,
allow one to crudely estimate the ionization fraction of hydrogen from the
observed \ntii\ and \nti\ column densities.
We find \ntii/(\ntii + \nti)~$\approx 0.15$ which suggests
$x \approx 15\%$, although N can be modestly overionized with respect to 
hydrogen because of its larger ionization cross-section.
We contend a similar effect explains the sub-solar
Ar$^0$/S$^+$ ratio observed for this predominantly neutral gas.

To perform a quantitative analysis of photoionization, we have performed
a series of calculations with the CLOUDY software package.  We caution
that the accuracy of these calculations are limited by a number of
uncertainties associated with the physical nature of the absorbing
gas and the incident ionization field.  
We consider here a series of single-phase, constant density models with
varying ionization parameter, the Haardt-Madau spectrum, and
assume $\N{HI} = 10^{20} \cm{-2}$ which is roughly consistent with the relative
strength of the low-ions and the total H\,I. 
In Figure~\ref{fig:FeFe}, we plot 
the predicted logarithmic abundance of low-ion X$_i$ 
relative to \feii\ relative to the intrinsic abundance of these two elements,
\begin{equation}
[{\rm X_i / Fe^+}] \, = \log (\N{X_i}/\N{Fe^+}) - \log({\rm X/Fe})_\odot \cmma
\label{eqn:xfe}
\end{equation}
against the \feiii/\feii\ ratio. 
We focus on \xfeii\ instead of \xhi\ because we 
are more concerned with the effects of photoionization
on interpretations of relative elemental abundances [X/Y] instead of 
metallicity [X/H] in the damped \lya systems.  
Departures of \xfeii\ from zero 
indicate that photoionization
corrections are required to calculate accurate elemental abundance ratios
from the low-ion column densities independent of dust depletion.
Positive (negative) \xfeii\ values would imply overestimates (underestimates)
of [X/Fe] from the low-ion ratios.
Finally, the black dashed line plots the ionization fraction,
$\log x$, on the same logarithmic scale.

At \feiii/\feii~$< -1.9$~dex, we note $\log x < -1$ indicating the hydrogen
gas is less than $10 \%$ ionized.
Therefore, we are confident that the gas at $v \approx 0 \mkms$ is
predominantly neutral.
In turn, we find the ionization corrections are
negligible for \siii, \crii, and \nkii\ relative to \feii\ and 
small for \suii, \alii, and \nti.  Only \ari\ requires a large ionization
correction which brings the corrected Ar/S ratio into closer
agreement with the solar abundance.
This gas resembles the $\beta$~Cen~A ISM sightline
analysed by \cite{sofia98} which shows similarly over-ionized Ar/S
in a gas with a relatively large neutral hydrogen fraction.
The fact that the inferred Ar/S ratio
remains sub-solar even after the ionization corrections implied by
Figure~\ref{fig:FeFe}
may reflect differential depletion but more likely points to an error in our
adopted ionization spectrum or possibly uncertainties in 
the atomic physics of Ar.
In either case, the difference reflects an error in our photoionization
modeling.

We emphasize that 
our conclusions on the ionization state of this gas are further supported
by the low \aliii/\alii\ and \feiii/\aliii\ ratios. 
Our results are also consistent with the observed \siiii/\siii, \siiv/\siii,
and \civ/\cii\ ratios.
Furthermore, we have also considered a Kurucz stellar ionizing spectrum 
with $T = 30000-50000$~K and more complicated two-phase scenarios
and have found no qualitative differences in these analyses.
In conclusion, the gas at $v \approx 0 \mkms$ has a low ionization
fraction $(x \ll 1)$ and ionization corrections to the observed low-ion
column densities are nearly
negligible except in the case of Ar$^0$.  Therefore, the chemical
abundances for this gas are accurately represented by the low-ion species.

\begin{figure*}[ht]
\begin{center}
\includegraphics[height=6.0in, width=4.5in,angle=-90]{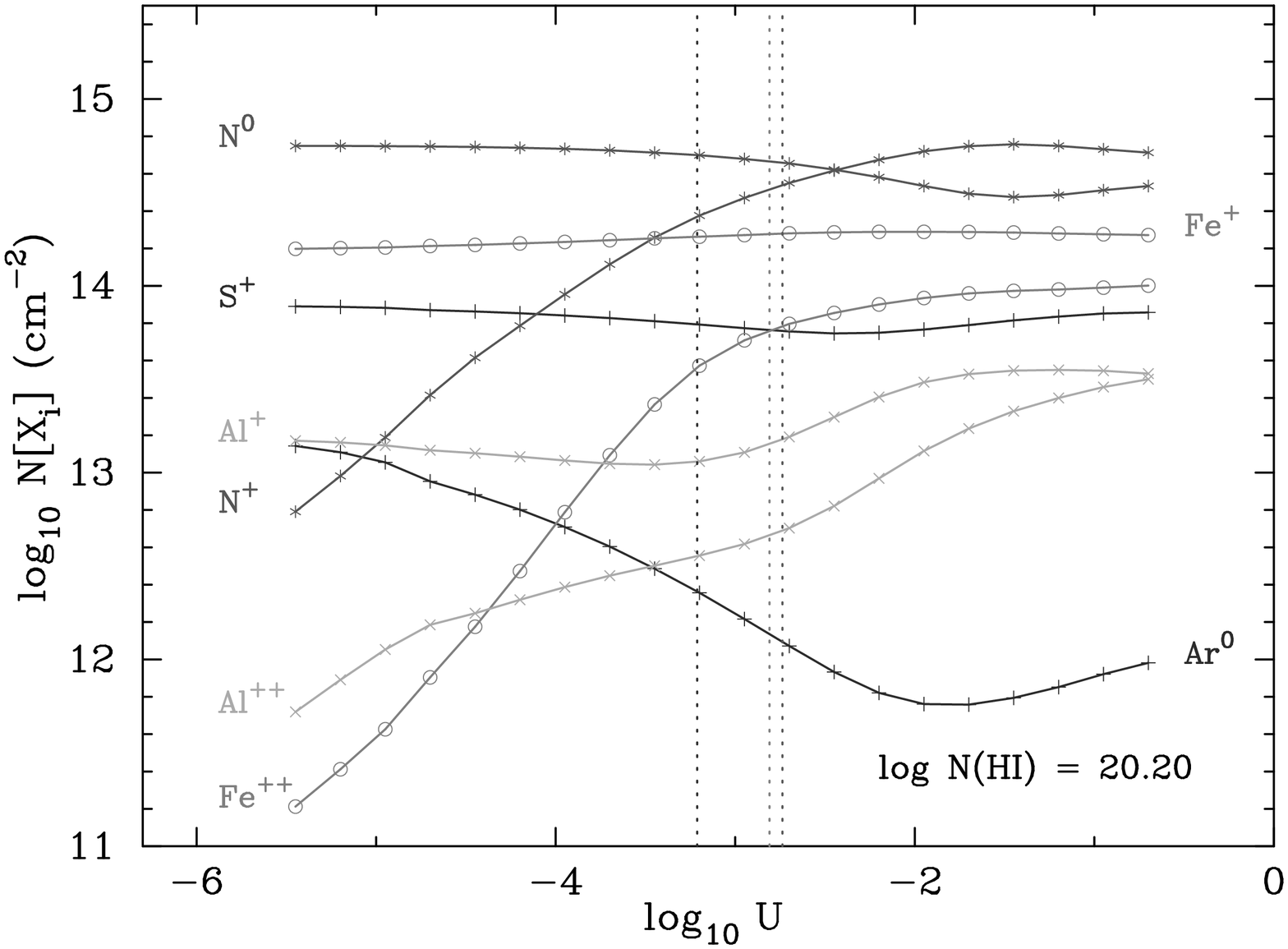}
\caption{
Predicted column densities $\N{X}$ for
a series of ionization parameters assuming $\N{HI} = 10^{20} \cm{-2}$,
a metallicity [Fe/H] = $-1$, and the Haardt-Madau EUVB spectrum at $z=2.5$.
The vertical dashed lines indicate the observed
ratios of \feiii/\feii, \ntii/\nti, and \ari/\suii\ 
and we note that the \aliii/\alii\ upper limit is consistent requires 
$\log U < -2$~dex.  We find that all four of these ionization
diagnostics are consistent with a single ionization phase and 
$\log U \approx -3$~dex.
}
\label{fig:U}
\end{center}
\end{figure*}

\subsection{$v \approx -26$ and $-47 \mkms$}

Now consider the gas at $v \approx -47$ and $-26 \mkms$ which exhibits
significantly larger \ntii/\nti, \aliii/\alii, and \feiii/\feii\ ratios
than the gas at $v \approx 0 \mkms$. 
In the case of \ntii\ and \feiii, we stress that these transitions lie in the
\lya forest and may suffer from contamination by coincident \lya clouds.
In the following, we proceed under the assumption that
such contamination is small.
Making a crude estimate of the ionization fraction 
from the observed \ntii/\nti\ ratio as above, we find $x \approx 50\%$ assuming
\ntii/\nti~$\approx -0.1$~dex.
We also note that the \aliii/\siii\ ratios are high relative to the
majority of damped \lya systems \citep{vladilo01}.  Altogether,
we expect this gas to be more highly ionized than the $v \approx 0 \mkms$ 
component.

The greatest challenge in accurately assessing the impact of photoionization
in the damped \lya systems is our ignorance of the physical nature of the
gas 'clouds' responsible for the absorption. 
We have no {\it a priori} knowledge of the metallicity, HI column density,
volume density, 
or the underlying nucleosynthetic or dust depletion patterns.
Furthermore, the absorption at any velocity could very
well be the superposition of many clouds each with a unique set of 
physical characteristics but coincident velocity along the sightline.
As such, there may be several clouds with different ionization states
giving rise to the various metal-line profiles.  
Our approach is to consider simplest models first and then
qualitatively assess the effects of more sophisticated scenarios.

\begin{figure*}[ht]
\begin{center}
\includegraphics[height=6.0in, width=4.5in,angle=-90]{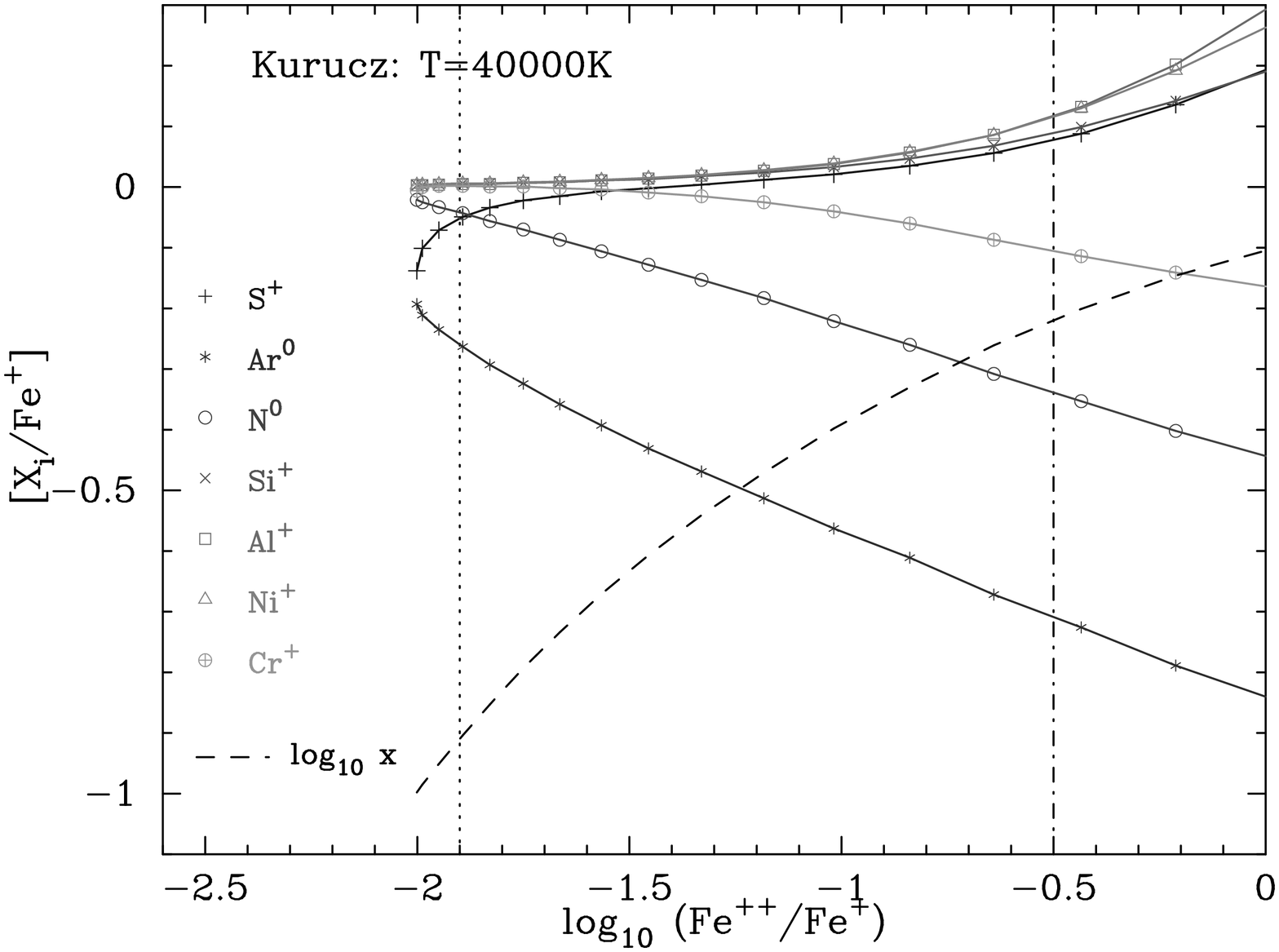}
\caption{
Predicted logarithmic abundance of low-ion X$_i$ 
relative to \feii\ relative to the intrinsic abundance of these two elements
against the \feiii/\feii\ ratio.   The analysis and presentation
is identical to Figure~\ref{fig:FeFe} except the ionizing spectrum is a Kurucz
$T = 40000$~K stellar model. 
}
\label{fig:FeFeKZ}
\end{center}
\end{figure*}

We consider the two following cases: (1) the gas arises in a single
cloud with one ionization phase; and (2) the gas arises in two gas
clouds with very different
degrees of ionization and possibly metallicity.  
Before proceeding, we present a line of
circumstantial evidence that the ionization corrections to the elemental
abundances might be small for this gas.   In Table~\ref{tab:summ},
we note that the
relative abundances of the low-ion species of S, N, Ar, Si, Fe, Ni, and
Cr are relatively uniform over all of these low-ion components.
In all cases, the ratios vary by less than 0.2~dex from component
to component.
If one adopts the unfounded assumption
that the gas in the entire damped \lya system shares similar elemental
abundances, then one would infer that the ionization corrections
are small.
Conversely, significant ionization corrections might require large
abundance variations, possibly from different chemical enrichment
histories or dust depletion patterns.  

\subsubsection{Single Phase Model}

The simplest scenario is that the gas 
is contained within a single, constant density cloud with one ionization 
phase\footnote{This treatment is identical to the scenario envisioned
by \cite{pro96} and the model designated H1 by \cite{vladilo01}.}.
If the cloud is subjected to an external radiation field, there 
will be a radial variation in the degree of ionization of the gas; 
the outer layers are highly ionized and the inner regions are
primarily neutral.
With the exception of \aliii\ \citep[e.g.][]{pro96}, 
the intermediate-ions arise in the
outer layers of the cloud because few high energy photons reach
the inner regions.
In this important respect, the single-phase
model mimics the two-phase scenarios presented next.
Because the \ntii/\nti, \feiii/\feii, and 
\feiii/\aliii\ ratios are similar for the gas at $v \approx -26$ and
$-47 \mkms$, we consider the two components together.  
The uncertainties in our photoionization modeling presumably exceed
any differences in the ionic column densities of the two velocity components.

Our analysis relies on simulations drawn from the CLOUDY software
package.
Figure~\ref{fig:U} presents the predicted column densities $\N{X}$ for
a series of ionization parameters assuming $\N{HI} = 10^{20} \cm{-2}$,
a metallicity [Fe/H] = $-1$, and the Haardt-Madau EUVB spectrum at $z=2.5$.
The H\,I column density and metallicity are reasonably consistent
with the total quantities inferred from the entire system.
We emphasize that at $\N{HI} \gtrsim 10^{20} \cm{-2}$, the exact
$\N{HI}$ value that one assumes has little impact on the single-phase
scenarios.
We also note that the $\N{X}$ values presented in Figure~\ref{fig:U}
are less important than the
relative values of the pairs of lines with identical color and symbols 
(e.g. \feiii/\feii).  The vertical dashed lines indicate the observed
ratios of \feiii/\feii, \ntii/\nti, and \ari/\suii\ 
and the \aliii/\alii\ upper limit requires 
$\log U < -2$~dex.   
We find that all four of these ionization
diagnostics are consistent with a single ionization phase where
$\log U \approx -3$~dex.
The agreement is not perfect, but it is acceptable given the number
of assumptions and uncertainties in the photoionization
modeling.  For example, a slightly softer ionizing spectrum would help
bring the predicted \ari/\suii\ ratio into better agreement with the
observations at $\log U \approx -3$~dex.  

The implications of the single phase model with $\log U \approx -3$
are as follows.  First, we derive an ionization
fraction $x \approx 50\%$ which indicates the gas is partially ionized. 
Second, if the metallicity of the velocity components at 
$v > -50 \mkms$ is constant, then the
$v \approx -26 \mkms$ component may exhibit a larger $\N{H}$ value than
the gas at $v \approx 0 \mkms$.
Third, the
degree of photoionization is significant enough to imply ionization
corrections to the relative abundances derived solely
from the low-ion species.   
Referring to Figure~\ref{fig:FeFe}, one notes that as \feiii/\feii~$\to 0$~dex
several of the low-ion ratios \xfeii\ require modest ionization corrections.
With the exception
of \ari, the corrections are generally small $(< 0.3$~dex) but 
significant compared to the typical statistical
errors in these measurements.
The vertical black dash-dot line in Figure~\ref{fig:FeFe} 
indicates the approximate observed
\feiii/\feii\ value of $-0.5$~dex.  For this value, the low-ion ratios
overestimate Si and Ni relative to Fe by $\approx 0.1$~dex
while Al, S, and N are underestimated by $\approx 0.2$~dex.
These ionization corrections are relatively small, but can have 
important implications for the relative elemental abundances of the damped \lya
systems.  For example, correcting the S and Si abundances gives
[S/Si] and [Si/Fe] values for the $v \approx -26 \mkms$ component
which differ by $\approx 0.3$~dex from the $v \approx 0 \mkms$ gas.
One concludes, therefore, that the relative elemental
abundances may differ significantly within this damped \lya system.

We have also considered a single-phase model with the Kurucz stellar
ionizing spectrum and a range of stellar effective temperatures.
These models imply similar corrections; compare against Figure~\ref{fig:FeFeKZ}
which plots a $T = 40000$~K model with $\N{HI} = 10^{20} \cm{-2}$.
The principle differences are that [\crii/\feii] shows a large offset
from solar and that [\suii/\feii] is positive at high \feiii/\feii\ values
and significantly negative at \feiii/\feii\ $< -2$~dex.
These differences arise from the softer Kurucz ionizing spectra
(e.g.\ a significant fraction of S is in S$^0$ at low ionization potential)
and they highlight the uncertainties in our models related to the assumed
input spectrum.  In theory, one might observe enough ions to discriminate
between various ionizing spectra but in practice one must allow for these
uncertainties. 

\begin{figure*}
\begin{center}
\includegraphics[height=6.0in, width=4.2in,angle=-90]{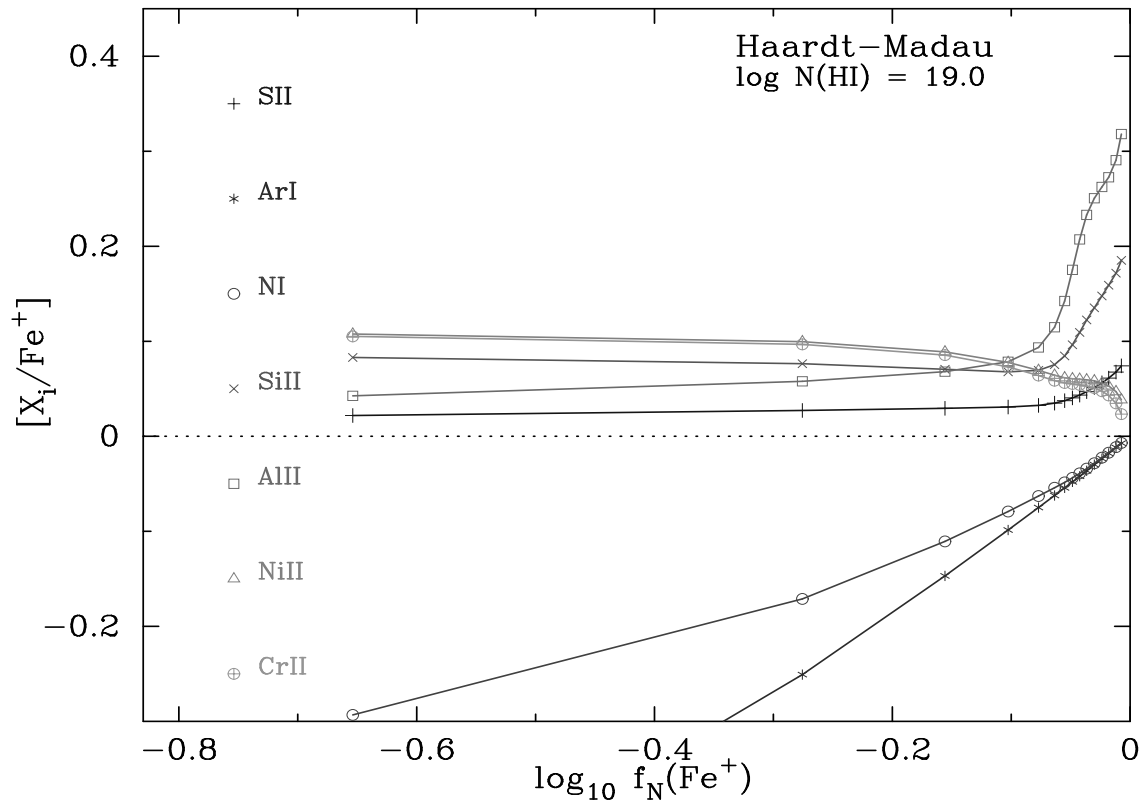}
\caption{
Offsets from the intrinsic abundances
of low-ion X$_i$ relative to \feii, \xfeii, against the fraction of
\feii\ arising in the neutral component \fN of our two-phase models.  
Similar to the single-phase models, the implied ionization corrections
are small for the majority of elements although there are several
quantitative differences.  
For this analysis, we have assumed the Haardt-Madau spectrum and that
$\N{HI} = 10^{19} \cm{-2}$ for the ionized component
}
\label{fig:fN}
\end{center}
\end{figure*}

\begin{figure*}
\begin{center}
\includegraphics[height=6.0in, width=4.2in,angle=-90]{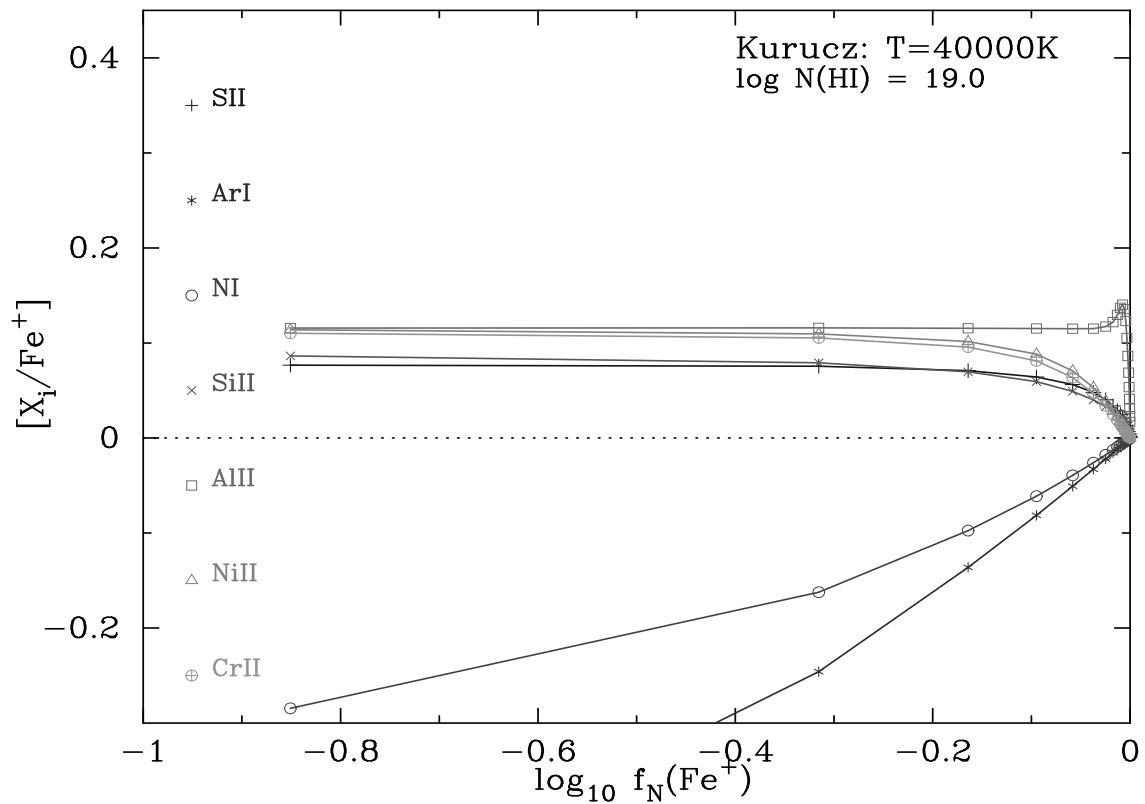}
\caption{Same as Figure~\ref{fig:fN} except we have assumed a
Kurucz ($T = 40000$~K) stellar model.
}
\label{fig:fNKZ}
\end{center}
\end{figure*}

\subsubsection{Two-Phase Model}

Although we found that the single-phase scenario gave reasonable agreement
to our set of photoionization diagnostics, we also wish to investigate
the effects of a two-phase scenario.  In particular, we want to determine
if the two approaches yield different estimates for the ionization
corrections \citep[e.g][]{vladilo01}.  
We will show that the two models predict similar
corrections but with important differences for \suii\ and \ari.

Consider the following scenario where the 
low and \\
intermediate-ions arise
in two parcels of gas with unique ionization
characteristics.  As discussed in $\S$~\ref{sub:intmgas}, 
it is possible that gas with properties similar to the ionized gas
at $v \approx -65 \mkms$ is entirely responsible for 
the intermediate-ions but also 
contributes a significant percentage of the low-ion column densities.
In fact, we estimated an ionized component could contribute
$\approx 20\%$ of the \feii\ gas observed at $v \approx -26 \mkms$.
This scenario resembles the physical conditions along the
sightline to HD~93521 in the Galactic ISM,  where \cite{sembach00}
describe the impact of the Warm Ionized Medium on measuring elemental
abundances.
For the damped system, a multi-phase interpretation is 
further supported by an observed offset between
the low and intermediate-ion gas at $v \approx -26 \mkms$.
Examining Tables \ref{tab:low} and \ref{tab:inter}, one notes that
the component which best describes the low-ion gas at $v \approx -26 \mkms$
is offset by $\approx 4 \mkms$ from the centroids of the
\aliii, \ntii, and \feiii\ component.
The difference is relatively small and may be the result
of a systematic error, but
we consider it more likely that the gas at 
$v \approx -26 \mkms$
is best described by at least two components with very different
ionization fractions\footnote{Unfortunately, the gas at 
$v \approx -47 \mkms$ is not precisely 
localized in velocity space and can be 
well described by a single or multiple clouds.}.
In any case, because the offset is small
one must consider the impact of the ionized gas on the observed low-ion
column densities and we treat the two clouds as
if they had identical velocities.
This treatment is similar to the scenario introduced by \cite{howk99} and 
extended by \cite{vladilo01}.

Consider the following model which assumes
the gas at $v \approx -26 \mkms$ is comprised of two components:
(i) a very neutral component (e.g. \feiii/\feii $\ll 1$) and;
(ii) an ionized component with identical
relative gas-phase abundances that contributes
all of the intermediate-ions observed at $v \approx -30 \mkms$. 
Even though the latter component is significantly ionized, it may
still exhibit significant column densities of several low-ion species,
(e.g.\ \suii, \siii, and \alii) and,  therefore, impact
the relative elemental abundances and metallicity derived from the low-ion
transitions.  

To illustrate the ionization corrections for this two-phase model we 
introduce a new parameter, \\
$f_N(X_i) \equiv (X_i)_{neutral}/(X_i)_{total}$, 
which is the fraction of low-ion
$X_i$ that is attributed to the neutral gas component. 
We have used CLOUDY calculations similar to those presented 
in the previous section
to calculate the ionization state of the ionized component but
assuming the HI column densities are lower than $10^{20} \cm{-2}$.
In this case the ionized component
contributes a small fraction of the total HI column density.
We constrain the two-phase model by requiring that it yields
a \feiii/\feii\ ratio of $-0.5$~dex.  Because this constraint refers
to a relative abundance instead of the absolute \feiii\ and \feii\ column
densities, our analysis is not very sensitive to the $\N{HI}$ value or
metallicity assumed for the ionized component.  Changing the $\N{HI}$ value
for the ionized component by
0.5~dex can be important, however, and we consider the effects below.

We first present results
assuming $\N{HI} = 10^{19} \cm{-2}$ in the ionized component and adopting
the Haardt-Madau spectrum.
Figure~\ref{fig:fN} plots the offset from the intrinsic abundances
of low-ion X$_i$ relative to \feii, \xfeii, against the fraction of
\feii\ arising in the neutral component \fN.  
As \fN\ increases in the
figure,  the ionized component contributes less \feii\ and therefore
there is a compensatory increase in the ionization state of this gas 
to match the observed \feiii/\feii\ ratio.
At low \fN\ values, one approaches the single-phase
model presented in the previous sub-section.  
Note that one derives a qualitatively similar set of
results by examining $f_N({\rm N^0})$ and constraining the predicted
\ntii/\nti\ ratio by the observed value.

Similar to the single-phase model, the implied ionization corrections
are small for the majority of elements although there are several
quantitative differences.  First, the offsets for [\suii/\feii]
are generally small or slightly positive whereas the single-phase
model predicted negative offsets approaching $-0.2$~dex.  
The difference can be attributed to the difference in optical depth
of the single-phase model and the ionized component in the two-phase
scenario where we have assumed $\N{HI} = 10^{20} \cm{-2}$
and $10^{19} \cm{-2}$ respectively.
In the former case, the self-shielding from H\,I is large enough
that S is much less ionized relative to Fe. 
Second, the
offsets for \ari/\feii\ are small at all but the lowest \fN\ values.
This is because we have assumed Ar is entirely in atomic form
in the neutral component.  If this two phase model is an accurate
representation of the gas, the neutral component must have a
sub-solar \ari/\suii\ ratio similar to the 
the neutral gas at $v \approx 0 \mkms$. 
Finally, we note large corrections for \alii, \siii, and \suii\ at
log~\fN~$> -0.07$ which corresponds to a very large ionization parameter
for the ionized component.  At very high levels of ionization, the
\siii, \suii, and \alii\ ions exhibit significantly greater abundances
relative to \feii.
We doubt, however, that the gas considered here has log~\fN~$> -0.07$.
As noted above, if the absorption at $v \approx -65 \mkms$ is representative
of the ionization component in our two-phase model, then
log~\fN~$< -0.1$.  Furthermore, it is very unlikely (if not nonphysical) to
assume that the neutral component contributes no \feiii\ gas.
Allowing some ionization of \feii\ to \feiii\ in the neutral
component drives the effective \fN\ value even lower.

We have also performed a two-phase analysis assuming a series of
Kurucz stellar spectra.  Figure~\ref{fig:fNKZ} presents the results
assuming $\N{HI} = 10^{19} \cm{-2}$, $T = 40000$~K, [Fe/H]~=~$-1$,
and \feiii/\feii~$= -0.5$~dex.
The results are qualitatively similar to those from the 
Haardt-Madau spectra (Figure~\ref{fig:fN}):
\siii, \alii, \suii, \crii, and \nkii\ all require small
negative ionization corrections relative to \feii\ while
\nti\ and \ari\ require large positive corrections.
In contrast with the Haardt-Madau results, all of the corrections
converge to 0~dex as \fN~$\to 1$.  For a stellar atmosphere with $T = 40000$~K,
one predicts significantly larger \feiii/\feii\ ratios than the 
Haardt-Madau spectrum, primarily due to the softer spectral slope of the
stellar flux. In fact, adopting a higher effective temperature 
(e.g.\ 50000~K) for the
stellar atmosphere brings the results into better qualitative agreement
with the Haardt-Madau models.
Finally, we have also considered different values of $\N{HI}$ in the
ionized component than $10^{19} \cm{-2}$.  
In general, lower $\N{HI}$ values lead to slightly
larger \xfeii\ values and higher $\N{HI}$ values imply smaller corrections.
The differences are typically smaller than 0.1~dex, however, and are
therefore on the level of the uncertainties inherent to our photoionization
modeling.

To summarize, the two-phase scenario implies small but important
corrections to the low-ion ratios.  
We favor this model over the single
phase model because of the likelihood that the 
intermediate and low-ion components at $v \approx -26 \mkms$ 
are not exactly aligned. 
Furthermore, we have identified ionized gas near the damped \lya system
(at $v \approx -65 \mkms$) which exhibit properties consistent with the
ionized component of our two-phase model.
Interestingly, 
the ionization corrections of the two-phase scenario
lead to relative abundances which are less conflicting
with the $v \approx 0 \mkms$ gas.
The challenge of the two phase scenario in comparison with the
single phase model, however, is that it introduces an additional
gas component with another set of uncertainties related to HI column
density, metallicity, geometry, etc.  These uncertainties prohibit
one from determining the \fN\ value and thereby precise ionization
corrections.   

\section{DISCUSSION}
\label{sec-discuss}

In the previous sections, we carefully considered the ionization state
of the gas comprising the damped \lya system at $z=2.62$ toward
GB1759+7539.  Qualitatively, we demonstrated that the gas in one major
velocity component is primarily neutral $(x<0.1$) while the 
remaining gas is partially ionized $(x \approx 0.5$).
In both cases, significant ionization corrections for \ari\ are required.
Furthermore, several elements in the ionized component -- notably N, S -- 
require small but important ionization corrections to infer accurate
elemental abundances from the low-ion column densities.
These corrections have important implications for our measurements
of the metallicity of this damped system and our interpretations of
the nucleosynthetic enrichment and dust depletion inferred from relative
chemical abundances.
Before proceeding, we wish to emphasize that {\it this system
exhibits a number of characteristics (e.g.\ variations among the low-ion
profiles) which separate it from the majority of damped systems.}
Therefore, one must be careful not to generalize the ionization properties
of this system to the entire sample of damped systems.  
At the same time, it is worth noting that the H\,I column density of this
system is significantly greater than the damped \lya threshold of
$2 \sci{20} \cm{-2}$ and would not have been expected to be partially 
ionized on the basis of its H\,I optical depth.

\begin{table}[ht]\footnotesize
\begin{center}
\caption{ {\sc
METALLICITY TABLE \label{tab:metal}}}
\begin{tabular}{lccccccc}
\tableline
\tableline
Model & $[$Fe/H$]$& $[$Si/H$]$& $[$S /H$]$ \\
\tableline
HM: No IC                               &$-1.26$&$-0.80$&$-0.81$\\  
HM: One-Phase                           &$-1.31$&$-0.90$&$-0.71$\\  
HM: Two-Phase, $f$(Fe$^+$)=--0.1        &$-1.31$&$-0.95$&$-0.88$\\  
HM: Two-Phase, $f$(Fe$^+$)=--0.5        &$-1.39$&$-0.97$&$-0.91$\\  
KZ: One-Phase                           &$-1.35$&$-0.92$&$-0.91$\\  
KZ: Two-Phase, $f$(Fe$^+$)=--0.1        &$-1.31$&$-0.95$&$-0.88$\\  
KZ: Two-Phase, $f$(Fe$^+$)=--0.5        &$-1.36$&$-0.97$&$-0.90$\\  
\tableline
\end{tabular}
\end{center}
\tablenotetext{}{These metallicity values are for the entire damped Lya system.  The photoionization 
corrections, however, have only been applied to the gas at $v \approx -26 \mkms$ and $-47 \mkms$.}
\end{table}

Consider first the impact of photoionization on the metallicity measurements
of this damped \lya system.  We focus on the metallicity of the neutral
hydrogen gas
because estimating the metallicity of all hydrogen gas (ionized + neutral)
would require assumptions on the distribution of HI as a function of 
velocity.  In addition, the metallicity of the neutral component
is more readily compared against theoretical models.
In Table~\ref{tab:metal}, we report a series of metallicity measurements
for S, Si, and Fe assuming a variety of photoionization scenarios for
the partially ionized gas\footnote{Note that the effects of photoionization
on the metallicity of this system are tempered by the fact that
only $\approx 50\%$ of the gas requires ionization corrections.}:
(i) no corrections; (ii) a single-phase cloud with the HM spectrum;
(iii,iv) a two-phase scenario using the HM spectrum
with two different assumptions on the level of ionization in the ionized component;
(v-vii) single and two-phase scenarios adopting the Kurucz spectrum 
with $T = 40000$~K.  
In the two-phase scenarios, the results are modestly 
sensitive to the $\N{HI}$ value
assumed for the ionized component.  For this discussion, we assume
$\N{HI} = 10^{19} \cm{-2}$. 

The results presented in Table~\ref{tab:metal} indicate
systematic offsets in the metallicity of this damped system
when one adopts ionization corrections. 
In general, however, the photoionization corrections imply a small,  
statistically insignificant decrease in the [X/H] values.
The $\lesssim 0.1$~dex corrections are on the
same order as the 0.1~dex uncertainty in the $\N{HI}$ value.
Therefore, even in the unlikely event that all of the damped \lya systems
require ionization corrections which are this large, 
it will have a relatively minor
impact on the chemical evolution history currently resolved by 
the damped \lya systems \citep[e.g.][]{pw00}.  The one possible exception,
however, is the chemical enrichment history 
derived from Zn$^+$ whose photoionization
balance is very poorly determined \citep{howk99}.

\begin{figure}[ht]
\begin{center}
\includegraphics[height=3.7in, width=2.8in,angle=-90]{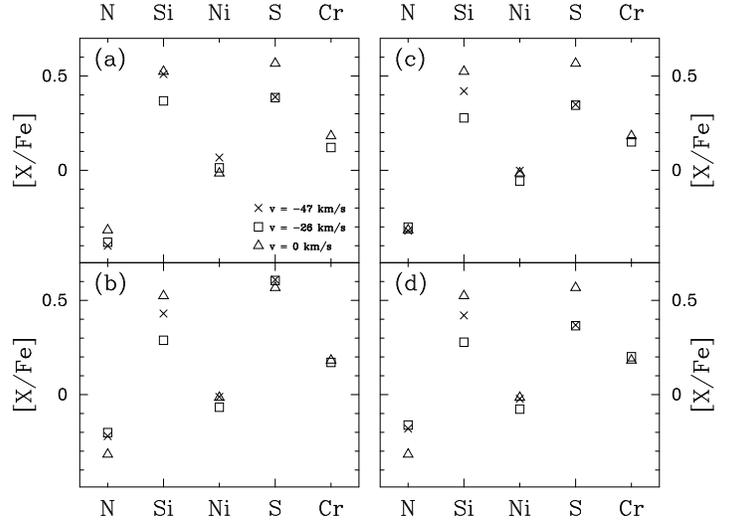}
\caption{
The relative abundance [X/Fe] of the three main velocity
components in this damped system for the Haardt-Madau input
spectrum.  We consider (a) no ionization
corrections; (b) a single-phase scenario; (c) a two-phase scenario with
\fN$= -0.3$; and (d) a two-phase scenario with \fN$= -0.1$~dex.
The $\times$'s mark the $v \approx -47 \mkms$ component, the
$\Delta$'s mark the $v \approx -26 \mkms$ component, and
the $\square$'s mark the $v \approx 0 \mkms$ component.
}
\label{fig:summHM}
\end{center}
\end{figure}

\begin{figure}[ht]
\begin{center}
\includegraphics[height=3.7in, width=2.8in,angle=-90]{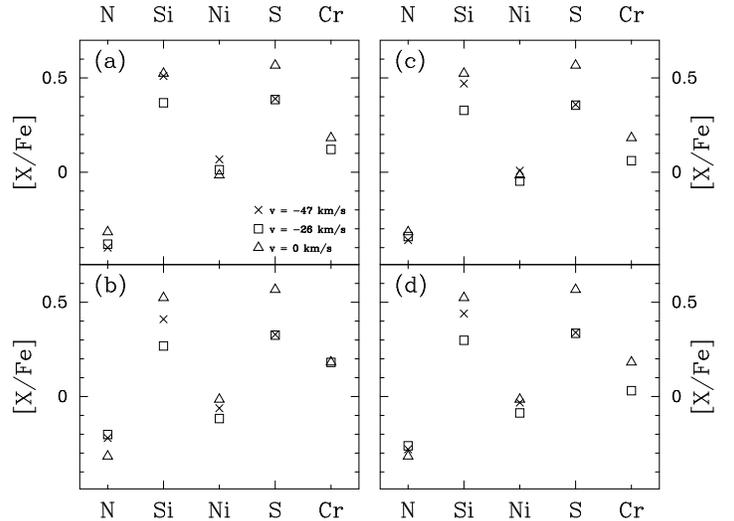}
\caption{Same as Figure~\ref{fig:summHM} but with the Kurucz input spectrum.
}
\label{fig:summKZ}
\end{center}
\end{figure}

In contrast to the metallicity measurements where factors of 
0.1 to 0.2~dex are generally unimportant, photoionization corrections
can significantly influence our interpretations of the relative elemental
abundances of the damped \lya systems.
A correction to \siii/\feii of 0.2~dex, for example, would imply
a very different dust-to-gas ratio and/or nucleosynthetic enrichment history
\citep[e.g.][]{pw02}.
This point is well illustrated in Figures~\ref{fig:summHM} and
\ref{fig:summKZ}
which present the relative abundance [X/Fe] of the three main velocity
components in this damped system for the Haardt-Madau and Kurucz input
spectrum respectively.  In each figure we consider (a) no ionization
corrections; (b) a single-phase scenario; (c) a two-phase scenario with
\fN\,$= -0.3$; and (d) a two-phase scenario with \fN\,$= -0.1$~dex.

Examining Figures~\ref{fig:summHM} and \ref{fig:summKZ}
we find that the
ratios of S/Fe and Si/Fe vary by $> 0.1$~dex among the various
ionization models.
Therefore, it is very difficult to assess the elemental abundances ratios of
the {\it ionized gas} in this damped \lya system.  
We are concerned that any damped system which is ionized to a similar
degree will prove just as challenging.
These uncertainties aside, we believe this system exhibits variations in
its relative abundances independent of photoionization.  These may be
due to differential depletion and/or varying nucleosynthetic enrichment
patterns.  In either case, this is one of the few examples of a 
damped \lya system where the abundances ratios are not uniform within
statistical error.  

The results of this paper highlight the importance of examining
photoionization diagnostics in a larger sample of damped \lya systems.
Although \aliii/\siii\ may provide a rough assessment of ionization
\citep{vladilo01},  a more precise analysis can be performed if 
one acquires observations of adjacent ions like \feii, \feiii\ and
\nti, \ntii.  Of course, the intermediate-ions require
observations with coverage deep within the \lya forest.
While it may be important to consider a full photoionization analysis
of the kind presented here for each damped \lya system, we believe the
following photoionization diagnostics (summarized in Table~\ref{tab:diag})
can provide a 'first-look' analysis
of the ionization state.  In all cases, our conclusions are derived
from the extensive set of CLOUDY photoionization models that we have
described throughout the paper.

\begin{itemize}

\item The \ari/(\siii,\,\suii) ratio is an excellent diagnostic for identifying
systems with low ionization factors, $x \approx 0$.  Specifically, 
a generic feature of photoionization models is that an 
observed value of [\ari/(\siii,\,\suii)]~$> -0.2$~dex requires $x < 0.1$.
For the 7 systems with Ar/Si measurements or limits \citep{molaro01,pw02}, 
they all suggest $x < 0.1$ except GB1759+7539.
It is also important to emphasize that a low  \ari/(\siii,\suii) ratio does
not require $x \gg 0$ given its large reionization cross-section
\citep{sofia98}.  In fact, we have argued that the gas at
$v \approx 0 \mkms$ in the GB1759+7539 damped system has $x < 0.1$ even
though it exhibits a significantly sub-solar \ari/\siii\ ratio.
Therefore, this diagnostic is only valuable for ruling out 
photoionization effects in a given damped system.
Finally, we should note that the \ari/(\siii,\suii) ratio could be affected
by nucleosynthesis and/or dust depletion.  On the other hand, Ar/S is solar
in systems with a wide range of metallicity in the local universe
\citep{henry99} and the majority of damped systems exhibit [S/Si]~$\approx 0$
indicating differential depletion will not modify the Ar, S, or Si abundances.

\item Unlike the \ari/(\siii,\,\suii) ratio, a determination 
of the \feiii/\feii\ ratio
can establish whether a given damped system is predominantly neutral
{\it or} ionized.  We find $x < 0.1$ in systems where 
\feiii/\feii~$< -1.6$~dex, and similarly, a system with \feiii/\feii~$> -1$~dex
will have $x > 0.5$.  A further advantage of this diagnostic is that the
\feiii/\feii\ ratio is independent of nucleosynthesis or dust depletion.
In short, an accurate measurement of \feiii\ via the Fe~III 1122 transition
is an extremely valuable probe of the ionization state of any quasar
absorption line system.

\item The \ntii/\nti\ ratio places similar constraints on $x$ as the
\feiii/\feii\ diagnostic.  We find $x < 0.1$ in systems showing
\ntii/\nti~$< -1$~dex and $x > 0.5$ in systems where
\ntii/\nti~$> -0.2$~dex.

\end{itemize}

\begin{table}[ht]\footnotesize
\begin{center}
\caption{ {\sc PHOTOIONIZATION DIAGNOSTICS\label{tab:diag}}}
\begin{tabular}{lccccccc}
\tableline
\tableline
Ratio & $x < 0.1$& $x > 0.5$& Comment \\
\tableline
\ari/(\siii,\suii) & $> -0.2$ & N/A & Sensitive to dust, nucleo \\
\feiii/\feii & $< -1.6$ & $> -1$ & \lya forest \\
\ntii/\nti & $< -1$ & $> -0.2$ & \lya forest \\
\tableline
\end{tabular}
\end{center}
\end{table}

In passing, we wish to comment on the implications of photoionization
for our kinematic modeling of the damped \lya systems \citep[e.g.][]{pro97}.
To assess the kinematic characteristics of an individual damped \lya 
system we choose a single, high S/N low-ion profile as representative
of the neutral gas for the entire system.
This approach was motivated by the fact that the low-ion profiles
track one another very closely \citep[e.g.][]{pro96}.  In the
case of the damped \lya system studied here, we have found significant
variations from low-ion to low-ion even prior to photoionization
corrections.  This variation will have a very small impact on the
measured velocity width ($\delv$; c.f.\ Prochaska \& Wolfe 1997)
of the damped \lya system, but the variations could impact the various
shape parameters implemented in our analysis.  Although this causes
us some concern, we emphasize that this system is unusual for exhibiting
a significant variation among the low-ion profiles.  In a future paper,
we will demonstrate the excellent chemical and ionization 
uniformity observed in the large
majority of damped \lya systems \citep{pro02}.

We also wish to reflect on the impact of photoionization regarding
the relationship between the observed \ciis\ profile and star formation
(see Wolfe et al.\ 2002 for a more complete discussion).
As noted in the introduction, the \ciis\ profile of the GB1759+75 damped
system does not closely
track the other low-ion profiles (Figure~\ref{fig:low}).
Specifically, the \ciis\ profile exhibits much greater optical depth
at $v \approx 0 \mkms$ than $v \approx -26 \mkms$.  We believe the 
gas at $v \approx 0 \mkms$ has properties similar to the Cold Neutral
Medium of the Galactic ISM \citep[e.g.][]{sav96}.  In contrast, 
if the gas at $v \approx -26 \mkms$ has physical properties similar to 
the Warm Neutral Medium or Warm Ionized Medium of the Galactic ISM,
then we would predict a low \ciis\ optical depth relative to the other
low-ions, as observed.  Therefore, we believe the components at
$v \approx 0$ and $-26 \mkms$ represent two distinct phases of gas.  
The net effect is that
it is difficult to assess the star formation rate within this
damped \lya system from its measured \ciis\ column density.

Finally, we wish to outline other avenues for investigating
the ionization state of the damped \lya systems.  
One might gain important insight into the properties of ionized
gas associated with damped systems by performing a comprehensive analysis
of the quasar absorption line systems with $\N{HI} \approx 10^{19.5} \cm{-2}$
\citep[e.g.][]{pro99,omeara01}.  These 'sub-DLAs' may show a wide range
of ionization states and would provide an excellent test of the modeling
implemented throughout this paper.  Furthermore, they are far more numerous
than the damped systems and one may design specific observational experiments
to focus on their ionization states 
(e.g.\ pre-select systems where \siiii, \ntii\, etc.\ readily observed).
Another more efficient approach would be to examine lower redshift 
systems where the \lya forest is a far less compromising effect.
The challenge here remains identifying enough systems toward very bright
quasars; a challenge which will be alleviated in part with the introduction
of COS on the Hubble Space Telescope.

\acknowledgments

The authors wish to extend special thanks to those of Hawaiian ancestry 
on whose sacred mountain we are privileged to be guests.  Without 
their generous hospitality, none of the observations presented 
herein would have been possible.
We wish to thank R. Carswell, F. Chaffee, and P. Outram for sharing their
spectrum of GB1759+75.
We acknowledge the very helpful Keck support staff for their efforts
in performing these observations. 
Finally, we thank Gary Ferland and CLOUDY associates for the CLOUDY
software package.
This work was partially supported by NASA through a Hubble Fellowship
grant HF-01142.01-A awarded by STScI to JXP.
JCH acknowledges support from NASA grant NAG5-10957 and from
NASA Long Term Space Astrophysics grant NAG5-3485 through the Johns
Hopkins University. 
The work of DT, NS, JMO, and DK was funded in part by 
grant NASA funds G-NASA/NAG5-3237 and NAG5-9224, and by
NSF grant AST-9900842.
AMW was partially supported by NSF grant AST 0071257.

\end{document}